\documentclass[superscriptaddress,11pt,amsmath,amssymb,floatfix]{revtex4-1}

\usepackage{graphicx,morefloats}
\usepackage{color,soul}
\usepackage{bm}
\usepackage{amssymb}
\usepackage{amsbsy}
\usepackage{amsmath}
\usepackage{amsthm}
\usepackage{times}
\usepackage{hyperref}
\usepackage{dsfont}
\usepackage{bbm}
\usepackage[normalem]{ulem}

\usepackage{hyperref}
\usepackage{graphicx}
\usepackage{xcolor}
\usepackage{multirow}

\usepackage{ccaption}% 

\usepackage{bm}

\makeatletter
\@ifundefined{date}{}{\date{}}
\makeatother

\usepackage{amsmath, amsfonts, amssymb}
\usepackage{mathtools}
\usepackage{physics}
\usepackage{bbm}
\usepackage{bm}
\usepackage{pifont} % dingbats

\usepackage{soul} % strikeout, etc.

\newcommand{\ncmd}{\newcommand}
\ncmd{\nn}{\nonumber}
\ncmd{\half}{\frac{1}{2}}
\ncmd{\mbf}[1]{\bs{#1}}
\ncmd{\gam}{\gamma}
\ncmd{\sig}{\sigma}
\ncmd{\pha}{\alpha}
\ncmd{\lam}{\lambda}
\ncmd{\dl}{\delta}
\ncmd{\kap}{\kappa}
\ncmd{\eps}{\epsilon}
\ncmd{\Lam}{\Lambda}
\ncmd{\Gam}{\Gamma}
\ncmd{\Dl}{\Delta}
\ncmd{\Ups}{\Upsilon}
\ncmd{\Om}{\Omega}
\ncmd{\om}{\omega}
\ncmd{\veps}{\varepsilon}
\ncmd{\vphi}{\varphi}
\ncmd{\vtheta}{\vartheta}
\ncmd{\tw}{\text{w}}
\ncmd{\pll}{\parallel}
\ncmd{\mc}{\mathcal}
\ncmd{\mf}{\mathfrak}
\ncmd{\bs}{\boldsymbol} % use this for "normal" compilations
\ncmd{\trans}[1]{{#1}^\intercal}
\ncmd{\eq}[1]{Eq. \eqref{#1}}
\ncmd{\fig}[1]{Fig. \ref{#1}}
\ncmd{\suppl}{\note{`Supplementary Information'}}
\ncmd{\bc}{\text{BC}}
\ncmd{\pd}[1]{\partial_{#1}}

\definecolor{blue2}{rgb}{0.2, 0.2, 0.6}
\definecolor{blue3}{rgb}{0.16, 0.32, 0.75}
\definecolor{darkred}{rgb}{0.8,0,0}
\definecolor{royalblue}{rgb}{0.0, 0.14, 0.4}
\definecolor{magenta}{cmyk}{0,.9,0,0.2}
\definecolor{amethyst}{rgb}{0.6, 0.4, 0.8}
\definecolor{cadmiumgreen}{rgb}{0.0, 0.42, 0.24}
\definecolor{deepcarmine}{rgb}{0.66, 0.13, 0.24}
\definecolor{forestgreen}{rgb}{0.13, 0.55, 0.13}

\ncmd{\new}[1]{{\color{darkred} #1}}

\ncmd{\para}[1]{\paragraph*{{\color{black}{\bf #1:}}} }

\ncmd{\note}[1]{{\color{gray}{[\ding{168} #1}]}}

\ncmd{\MMnote}[1]{{\color{blue}{[\ding{168} {\bf #1}}]}}
\ncmd{\mmc}[1]{{\color{blue}{[MM: #1}]}}
\ncmd{\YWnote}[1]{{\color{purple}{[\ding{168} {\bf #1}}]}}

\ncmd{\sur}[1]{{\color{forestgreen}{ #1}}}

\ncmd{\qs}[1]{{\color{magenta}{ #1}}}

\ncmd{\qsnote}[1]{{\color{red}{ #1}}}
\ncmd{\yw}[1]{{\color{blue2}{#1}}}

\begin{document}
\hyphenation{va-ni-sh-ing}
\begin{center}
\thispagestyle{empty}

{\large\bf Amplified response of cavity-coupled quantum-critical systems}
\\ [0.3cm]

Shouvik\ Sur$^{1,\dagger}$,
Yiming\ Wang$^{1,\dagger}$,
Mounica\ Mahankali$^{1}$,
Silke\ Paschen$^{2,1}$,
Qimiao\ Si$^{1,\ast}$ 
\\[0.3cm]

$^1$Department of Physics and Astronomy, Extreme Quantum Materials Alliance,
Rice Laboratory for Emergent Magnetic Materials, Smalley-Curl Institute, Rice University, Houston, Texas 77005, USA\\[-0.cm]

$^2$Institute of Solid State Physics, Vienna University of Technology, Wiedner Hauptstr. 8-10, 1040
Vienna, Austria

\end{center}

\vspace{0.16cm}

\noindent
\textbf{ 
A quantum critical point develops when matter undergoes a continuous transformation between distinct ground states at absolute zero. It hosts pronounced quantum fluctuations, which render the system highly susceptible to external perturbations. While light-matter coupling has rapidly moved forward as a means to probe and control quantum materials, the capacity of quantum critical fluctuations in the photon-mediated responses has been largely unexplored.
Here we advance the notion that directly coupling a quantum critical mode to a quantized cavity field dramatically facilitates the realization of the elusive superradiant phase transition  
in equilibrium, circumventing at once the key obstacles that have prevented its attainment in spite of decades of pursuit. 
The superradiant phase transition develops 
far below the ultrastrong regime of light-matter couplings, 
and the transition is accompanied by 
the hybrid system showing 
strongly enhanced intrinsic squeezing and amplified quantum Fisher information.
We also identify candidate cavity quantum materials platforms for validating the proposed effect.
Our findings suggest a general principle by which quantum criticality amplifies
the response to cavity photons. They also demonstrate 
that cavity coupling accesses
the elevated quantum entanglement of the underlying matter at quantum criticality, thereby pointing to a pathway
towards realizing the potential of highly collective quantum materials to expand the capacities of quantum information science.
} 

%\\ 
\medskip
\medskip
{
\noindent $^\dagger$ These authors contributed equally: Shouvik Sur and Yiming Wang.

\noindent $^\ast$ Corresponding author: \href{qmsi@rice.edu}{qmsi@rice.edu}
}
\clearpage

\newpage

\clearpage

\newpage

\vskip 0.1 in
\noindent
{\large \bf{Introduction} } 

Strong correlations give rise to a rich variety of unusual physical properties \cite{Keimer2017, PaschenSi}.
This is especially so for systems in a quantum critical regime, where quantum fluctuations are  
pronounced and physical responses are 
enhanced~\cite{PaschenSi,QC-proc,lohneysen2010special,Sachdevboo,Coleman-Schofield,HuChenSi}.
To understand the highly collective quantum critical fluids,
new means of probing them are highly desired.
A defining characteristic of quantum criticality is the mixing of statics and 
dynamics~\cite{Hertz,Chakravarty}, 
and indeed, singular dynamical responses 
have been demonstrated in the 
quantum critical regime~\cite{Aronson, Schroder, ProchaskaSingularCharge2020, Fang2025, Mazza2024,kirchner-rev}.
They not only corroborate the existence of the underlying {quantum critical point (QCP)},
but also 
characterize the nature of the quantum criticality~\cite{HuChenSi}.
As such, external dynamical perturbations are capable of elucidating
the quantum critical state~\cite{Kroha}.

Here, we address how coupling to an optical cavity provides an important new means of exploring the amplified responses of quantum criticality {at thermodynamic equilibrium}.
In a larger context, our approach is motivated by the increasing recognition that light-matter coupling can effectively interrogate and manipulate quantum materials~\cite{schlawin2022cavity, basov2025polaritonic, jarc2023cavity, keren2025cavity}, and engineer novel quantum states at thermodynamic equilibrium that are part light and part matter~\cite{lu2025cavity}. 
More specifically, a 
cavity introduces a single mode of quantized electromagnetic radiation that can be coupled to matter degrees of freedom through dipolar or Zeeman type interactions~\cite{knight1978super}.
It has been studied extensively in the 
pursuit of a superradiant  phase~\cite{carusotto2013quantum, forn-review}.
The latter is characterized by a macroscopic occupation of a cavity-photonic  
mode~\cite{hepp1973superradiant,wang1973}. 
The Dicke model,
which describes a collection of two-level subsystems interacting with a quantized cavity mode, provides the standard setting for exploring superradiant phase transition (SRPT)~\cite{Dicke1954, garraway2011dicke,  kirton2019introduction}. 
The transition, taking place in the thermodynamic limit, 
is characterized by the development of a  macroscopic occupation of the cavity mode and a spontaneous collective polarization in the matter sector.
In the absence of detuning, the SRPT  requires a  light-matter  coupling strength on the order of  or exceeding $10\%$ of  the cavity mode's energy, placing it in the ultrastrong coupling regime~\cite{forn-review, frisk2019ultrastrong}.
{Moreover, a ``no-go'' theorem, resulting from the requirement of gauge invariance~\cite{rzazewski1975phase, nataf2010no}, restricts the ability of dipolar couplings in inducing an SRPT at thermodynamic equilibrium~\cite{kirton2019introduction}.}
As a manifestation of these 
two challenges, the
experimental realization 
of equilibrium SRPT in cavity-coupled systems remains elusive in spite of decades of efforts~\cite{frisk2019ultrastrong,kim2025observation}.
Therefore, identifying mechanisms that ease access  to the superradiant  phase  and the concomitant SRPT  is of broad interest, as it enables controlled studies of collective quantum phenomena in light-matter interacting systems.

\begin{figure}[h!]
 \centering
\includegraphics[width=.95\columnwidth]{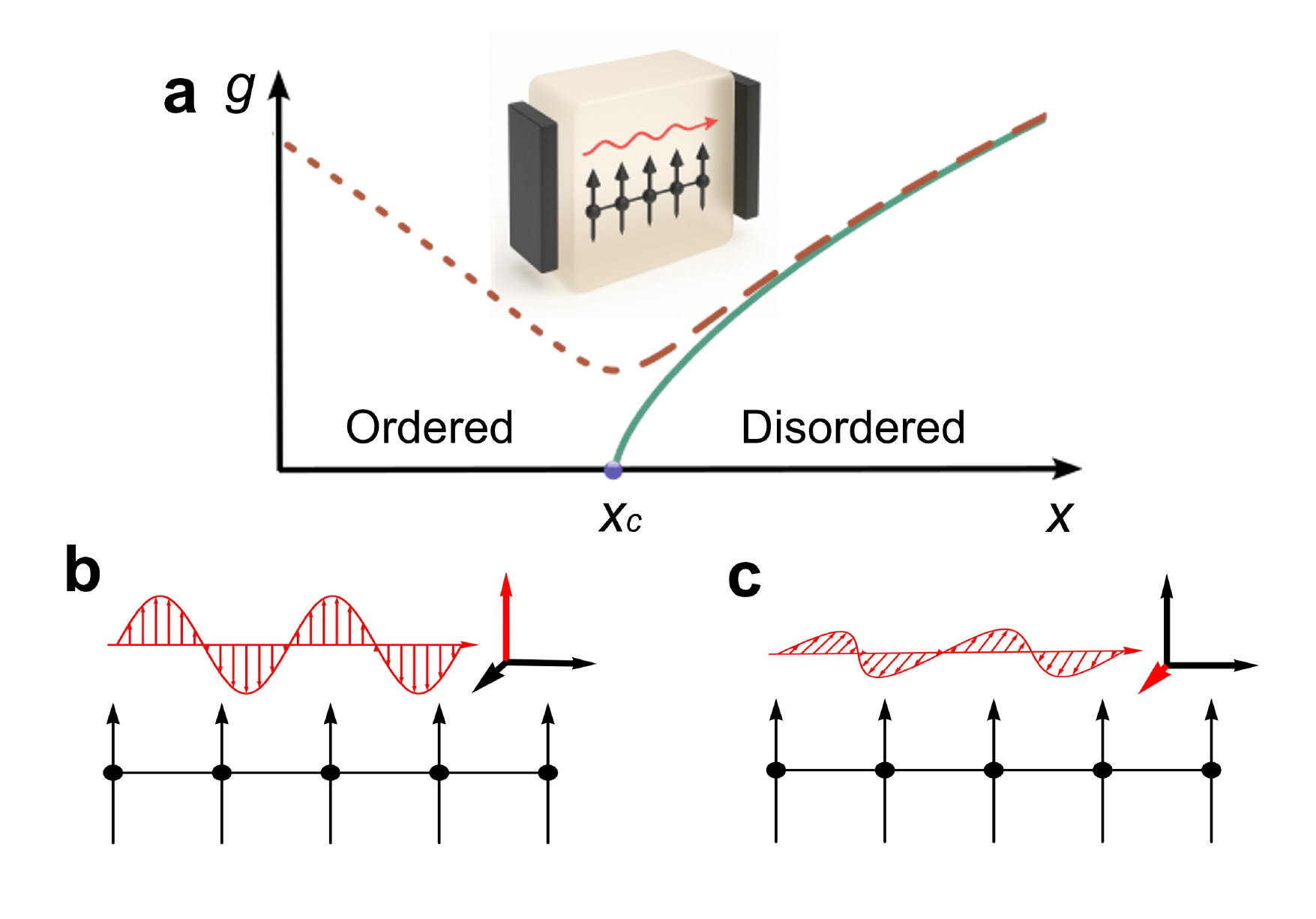}
\caption{
{\bf Schematic of a cavity mode coupled to matter degrees of freedom near a {QCP}. ~~~} {\bf a},
Phase diagram as a function of a non-thermal tuning parameter $x$. The 
coupling strength ($g_c$) required to induce a superradiant quantum phase transition (SRPT) is minimized at the matter QCP ($x_c$). The minimum $g_c$ vanishes when the cavity model couples directly to the critical mode (solid curve) but remains nonzero otherwise (dashed curve).
{\bf b}, Schematic of the cavity magnetic field aligned parallel to the Ising spin-coupling direction. The red sinusoidal curve with arrows represents the cavity magnetic field mode.
{\bf c}, Same setup as in {\bf b}, but with the cavity magnetic field oriented perpendicular to the Ising spin-coupling direction.}
\label{fig:sche}
\end{figure}

We focus on the effect of cavity coupling in a canonical {quantum} magnetic  
system across its QCP, as illustrated in Fig.~\ref{fig:sche}{\bf a}.
Importantly, when the cavity mode 
directly ({ i.e.} bilinearly, {through a Zeeman
coupling that avoids the no-go theorem}, { cf.} Fig.~\ref{fig:sche}{\bf b})
 couples  to the degree of freedom that 
exhibits quantum critical fluctuations,
we show here that
SRPTs can be realized  
{far below the ultrastrong coupling limit of light-matter interactions} (Fig.~\ref{fig:ising0}{\bf a},{\bf c}); thus, our work bypasses both challenges in the long-standing pursuit of equilibrium SRPT. 
The superradiant states, thus obtained, are highly squeezable in the vicinity of the SRPTs, and support a large multipartite entanglement that can be witnessed by the quantum Fisher information.
Since both aspects provide  valuable metrological resources, our work indicates  cavity quantum materials tuned to the vicinity of matter-QCP can {potentially} serve as particularly efficient quantum sensors.
By analyzing the scaling behavior of intrinsic squeezing close to the  SRPTs, 
we show that in such systems,
the coherent mixing of critical
matter modes and cavity photons generates a superradiant state that can be squeezed more efficiently than that in the original Dicke model.
{To compare and contrast our results with those for the Dicke model, we restrict our consideration to models of quantum magnetism with spin $1/2$.}
That a direct coupling of the cavity mode to a quantum critical degree of freedom   
enables the underlying matter quantum criticality to sharply amplify the responses to  cavity coupling represents a key new insight, which has not been recognized in previous work on such cavity-coupled 
systems~\cite{lee2004, gammelmark2011phase, zhang2014quantum, rohn2020ising, puel2024, langheld2024quantum,zhu2019dicke, mendoncca2025role, rao2025unilateral}.
{Conversely, our results show that
cavity coupling provides a means to access
the enhanced quantum entanglement of the underlying matter at quantum criticality, thereby suggesting the potential of highly collective quantum materials for 
 quantum information science. Finally,}
%Further,
we identify specific 
materials that are 
amenable to cavity materials engineering~\cite{lu2025cavity} and provide concrete platforms for 
validating the effect we have advanced.

\vskip 0.1 in
\noindent
{\large \bf{Results} }

\noindent {\bf{Cavity-coupled quantum critical system.~~}}
A quantum spin system coupled to cavity photons contains the following  ingredients.
The cavity mode, denoted by 
the field operator $\hat a$, has frequency $\omega_0$. 
It couples to the $\alpha$-th component ({ c.f.} 
Figs.~\ref{fig:sche}{\bf b},{\bf c}) of the magnetization of the quantum spin system (${\hat S}^{\alpha}_{\mathbf r}$), with a {collective} coupling constant $g$.
The matter sector is described by the Hamiltonian $H_{\text{spin}}$, 
in various spatial dimensions.
The overall Hamiltonian of the light-matter coupled system takes the following form:
\begin{align}
\hat H =& \omega_0 \hat a^\dagger \hat a 
+ \frac{g}{\sqrt{N}}(\hat a + \hat a^\dagger) \sum_{\mathbf r} \mathbf n \cdot \mathbf  {\hat S}_{\mathbf r} +  \hat H_{\text{spin}} \, ,
\label{eq:fullH}
\end{align}
where $\mathbf n$ is a unit vector 
controlling the spin projection 
that couples with the photon~\cite{emary2004},
{and $N$ is the number of spin sites.}
Here, the magnetic component of light Zeeman-couples to the localized spins of the quantum spin system. 
Consequently, $\hat H$ is gauge invariant (see Methods).

For concreteness, we will primarily focus on the ferromagnetic transverse field Ising model (TFIM), 
\begin{align}
\hat H_{\text{spin}} =  -J \sum_{\expval{\mathbf r, \mathbf r'}} \hat S_{\mathbf r}^x \hat S_{\mathbf r'}^x   - h\sum_{\mathbf r}  {\hat S}^{z}_{\mathbf r} \, ,
\label{eq:model}
\end{align}
where $J\geq 0$ 
describes the strength of the Ising spin-spin interactions and $h$ is 
a transverse field that also 
specifies the detuning in the cavity. 

\noindent {\bf{Coupling to a critical degree of freedom.~~}}
Our primary focus will be on the cavity photons that are Zeeman-coupled to the 
order parameter of the underlying Ising quantum phase transition 
(Fig.~\ref{fig:sche}{\bf b}).
In this case, the cavity mode 
bilinearly couples to the order parameter and, as such, the singular quantum critical fluctuations of the order parameter 
directly affect the response of the photon field.

This corresponds to the choice $\mathbf n = \hat x$, so that the photon field is linearly coupled 
to the magnetization $\hat M_x$. 
In the absence of the cavity coupling, the system undergoes a continuous ferromagnetic quantum phase
transition across the critical field, $h = h_{\text{TFIM}}$ (which equals to $J$ in the large-$S$ limit),
as the transverse field $h$ is tuned for a fixed Ising exchange interaction $J$.
In the ferromagnetic phase, the order parameter---the net magnetization $m_x = \expval{\hat M_x}$---is nonzero. 
In the paramagnetic phase, the order paramater vanishes.
The static magnetic susceptibility, $\chi_x$, diverges upon tuning
$h$ across the critical field, $h_{\text{TFIM}}$  (see Methods).

In this case, the ferromagnetic and the superradiant phases mutually cooperate because both the light-matter ($g$) and Ising ($J$) terms  weaken the field polarized state while commuting with each other. 
{Since in the $\mathbf n = \hat x$ limit the model is not exactly solvable, in} order to demonstrate this cooperation and explore its consequences, we will {first} obtain the zero-temperature phase diagram supported by $\hat H$ in the large-$S$  limit, and, subsequently,
verify these predictions for $d=1$
{in the spin-$1/2$ case} through density matrix renormalization group (DMRG) calculations. 

We isolate the $\mathbf k = 0$ magnon mode (henceforth, represented by $\hat{\mf b}_0$; see Methods), and solve for the polaritonic normal modes.
The vanishing of the dispersion at a critical coupling, $g_c$,
triggers a Bose-Einstein condensation (BEC) in the corresponding polaritonic mode,
which amounts to an SRPT.
In the standard Dicke model, corresponding to
$(J, S) \to (0, 1/2)$,
$g_c^2 = \omega_0 h$, 
which indicates the need for an ultrastrong light-matter coupling at weak detunings~\cite{forn-review}. 
A nonzero $J$ 
reduces $g_c$ and favors the nucleation of a superradiant state. 

A key result of our work is that the critical cavity-spin coupling for the SRPT vanishes 
at the TFIM QCP. 
Consider a fixed $J$, as $h \to h_{\text{TFIM}}^+$,
$g_c$ vanishes (as seen from {computations in the large-$S$ limit in} Methods, Eq.~\ref{eq:gc}).
For a fixed $J$ and $\omega_0$, superradiant states are present in the entire region bounded 
from below by the curve  $ g = g_c(h/J) \Theta\qty((h/J) - (h/J)_{\text{TFIM}})$ on the $(h, g)$ plane.
In Fig.~\ref{fig:ising0}{\bf a} we identify this region by plotting $\langle a \rangle$ 
 for the one-dimensional 
cavity-TFIM.
As the phase boundary is approached from the $g > g_c$ side,  $\langle a \rangle$  vanishes continuously as $g$, $g^2$, and $\sqrt{g - g_c}$ for $h < h_{\text{TFIM}}$, $h = h_{\text{TFIM}}$, and $h > h_{\text{TFIM}}$, respectively, as depicted in Fig.~\ref{fig:ising0}{\bf b} and described in detail in {Supplementary Note 1}.
This variation in the scaling of $\langle a \rangle$ indicates the presence of distinct scaling regimes in the superradiant phase that reflect the phase diagram of the underlying 
matter sector.

An important question is what happens in the extreme quantum limit. 
To address this issue,
we have performed DMRG simulations for  spin-$\half$ {(i.e. $S = \half$)} TFIM coupled to a cavity mode (see {Supplementary Note 2} for details).
Figs.~\ref{fig:ising0}{\bf c},{\bf d} show 
a line of continuous quantum phase transitions between the  Ising-paramagnetic normal phase and a superradiant phase on the $h > h_{\text{TFIM}}$
side of the phase diagram.
The numerically obtained phase boundary is such that $g_c \propto (h - h_{\text{TFIM}})^\zeta$ with $\zeta = 0.65 \approx 2/3$. Not surprisingly, the scaling exponents obtained by DMRG simulations deviate from the large-$S$ result.
Importantly, though, our result shows that the phase diagram is robust when the quantum fluctuations are fully accounted for.

\noindent {\bf{Intrinsic squeezing and quantum entanglement.~~}}
The extreme propensity of the fully quantum system towards an SRPT, as revealed by our
DMRG simulations, sets the stage for us to determine the metrological and quantum entanglement implications of our findings. 
The intermixing between the cavity mode and the critical spin degree of freedom captures the coherence between the light and matter sectors, which is described in terms of
an intrinsic two-mode squeezing~\cite{simon2000peres,duan2000inseparability,braunstein2005quantum}.
Specifically, {in the large-$S$ limit,} the variance of the polaritonic operator, 
\begin{align}
\hat X_{\mathrm{\theta, \phi, \psi}}(h/J) = \half [e^{i \phi}(\cos\theta ~ \delta \hat a + e^{i\psi} \sin\theta ~ \delta \hat{b}) + \mbox{h.c.}] \, ,
\label{eq:X}
\end{align}
with $\delta \hat a$ and $\delta \hat b$ representing fluctuations about   $\expval{\hat a}$ and $\expval{\hat{\mf b}_0}$, respectively, and $(\phi, \psi, \theta)$ being optimization parameters, is minimized to zero at the SRPT (see Methods).
The intrinsic squeezing in the limit $g \to g_c^+(h/J)$ is 
sensitive to the three superradiant regimes identified above, and the  minimum variance, $\Delta X_{\text{min}}^2(h/J)$, scales as $(g - g_c)^{0}$, $(g - g_c)$, and $(g - g_c)^{1/2}$ for $h < h_{\text{TFIM}}$, $h=h_{\text{TFIM}}$, and $h>h_{\text{TFIM}}$, respectively; the last two cases for $S=1/2$ are shown in Fig.\ref{fig:ising0}{\bf e}.
We note that the vanishing of $\Delta X_{\text{min}}^2(h/J)$ with $g \to g_c$ for $h>h_{\text{TFIM}}$ can be alternatively viewed as the existence of a perfect intrinsic squeezing at a fixed $g$  as $h$ tunes the system across an SRPT (c.f. Fig.~\ref{fig:ising0}a). This is a remarkable outcome when contrasted with the weak squeezing found in pure TFIM ($g = 0$) as $h$ is tuned across $h_\mathrm{TFIM}$~\cite{frerot2018}.

Importantly, in the vicinity of the SRPT, the squeezing is stronger at the QCP compared to the case of pure Dicke model.
In particular, comparing what happens at the QCP ($h=h_{\text{TFIM}}$) with that in the disordered regime ($h>h_{\text{TFIM}})$, 
for a fixed distance from the respective SRPTs, $\delta g \equiv g - g_c$,  
$\Delta X_{\text{min}}^2(h= h_{\text{TFIM}}) / \Delta X_{\text{min}}^2(h>h_{\text{TFIM}}) \sim \sqrt{\delta g}$,
which vanishes as $\delta g \to 0$.
This reflects the interplay between the approach to the SRPT and the underlying quantum criticality of the TFIM.
The 
reduction of $\Delta X_{\text{min}}^2$ at the QCP from that in the disordered regime
reflects the increased precision with which 
$\hat X_{\text{min}}$ can be measured {in principle} at the QCP.
By contrast, for the ordered regime ($h < h_{\text{TFIM}}$) and at sufficiently weak $g$, 
the spin-sector possesses a long range order, which is not conducive to squeezing; 
here the only meaningfully squeezable quadrature comes solely from the photon sector, which does not exhibit a perfect squeezing~\cite{emary2003chaos} (see {Supplementary Note 3}).

The elevated intrinsic squeezing 
at
the QCP ($h = h_{\text{TFIM}}$)
indicates the enhancement of light-matter quantum entanglement~\cite{duan2000inseparability}. 
The latter can be 
described in terms of the variance of the variable conjugate to $\hat X_{\text{min}}$. 
The procedure for identifying this conjugate variable, $\hat X_{\text{max}}$, is presented in the Methods.
The variance, $\Delta X_{\text{max}}^2$,
becomes large, as shown in Fig.~\ref{fig:ising0}{\bf f};
near the QCP, $\Delta X_{\text{max}}^2$ diverges $\sim \frac{1}{g-g_{\mathrm{c}}}$, which is stronger than
the $\sim \frac{1}{(g-g_{\mathrm{c}})^{1/2}}$ form arising
in the Dicke model as well as in the disordered regime ($h> h_{\text{TFIM}}$).
For the pure state we are considering, this variance is proportional to (is equal to $1/4$ of) 
the polaritonic quantum Fisher information~\cite{fadel2024quantum}, capturing the degree of light-matter quantum entanglement.

\noindent {\bf{Cavity coupling to a non-critical mode.~~}}
For comparison, we now turn to the case where the light-matter coupling is orthogonal to the Ising order parameter, corresponding to $\mathbf n = \hat y$ 
({ i.e.} Fig.~\ref{fig:sche}{\bf c}).
In this  case, the light-matter and Ising terms no longer commute.
Consequently, the ferromagnetism competes with the superradiant state, and their respective fluctuations mutually frustrate each other.
This competition results in a complex phase diagram~\cite{lee2004,gammelmark2011phase} as shown in Fig.~\ref{fig:ising}{\bf a}.  
The SRPT boundary reaches a  minimum in the vicinity of the TFIM QCP, which underscores the role of the matter QCP in facilitating the superradiant phase
{(c.f., Fig.~\ref{fig:sche}{\bf a})}. 
Moreover, this minimum corresponds to a tricritical point that generates an anomalous scaling for $\expval{\hat a} \sim (g - g_c)^\beta$ with $\beta \approx 0.25$ in its vicinity and supports a rich set of crossover behaviors, as portrayed in Fig.~\ref{fig:ising}{\bf b} (also, see  Methods and {Supplementary Note 4}).

\noindent {\bf{Other models and robustness.~~}}
We now address the robustness of the SRPT facilitated by the 
matter quantum criticality by considering a different cavity-coupled model with the spin Hamitlonian describing 
the 1D ferromagnetic XY model, 
\begin{align}
\hat H_{\text{spin}} = - \frac{J}{2} \sum_{{i}} \qty[ (1 + \Delta) \hat S_{i}^x \hat S_{i+1}^x + (1 - \Delta) \hat S_{i}^y \hat S_{i+1}^y ] \, .
\label{eq:XY}
\end{align}

Since $\hat H_{\text{spin}}$ supports distinct types of orderings (see Methods), a fixed light-matter vertex can represent coupling to either critical or non-critical matter-modes, depending on the location of the model parameters in its phase diagram. 
Here, the choice $\mathbf n = \hat x$ or $\hat y$ ($\mathbf n = \hat z$) in Eq.~\eqref{eq:fullH} corresponds to coupling the cavity mode to a critical (non-critical) matter mode.
The main contrast with the TFIM lies in the fact that as $\Delta \to 0^-$ ($\Delta \to 0^+$), for $\mathbf n = \hat x$ ($\mathbf n = \hat y$), the diverging correlation length of the fluctuations in the $\hat S^x$ ($\hat S^y$) channel  continuously suppresses the $g_c$, even though magnetic order persists in the $\hat S^y$ ($\hat S^x$) channel (see Sec.V of the SI). 
For any choice of $\mathbf n$, however, $g_c$ is minimized in the vicinity of $\Delta = 0$ ({c.f. Supplementary Note 5}), consistent with our earlier analysis of the  cavity-TFIM variants.

\vskip 0.1 in
\noindent
{\large \bf{Discussion} } 

\noindent {\bf{Experimental implications. ~~}}
The ferromagnetic-TFIM quantum phase transition
can be studied in the 
quasi-one dimensional 
materials CoNb$_2$O$_6$~\cite{coldea2010quantum}, 
as well as higher dimensional systems, such as LiHoF$_4$~\cite{bitko1996quantum} and CrI$_3$~\cite{huang2017layer}.
These materials can be directly coupled to a quantized cavity mode to access the 
propensity for SRPT, and elevated photon-matter squeezing and entanglement  in the vicinity of the magnetic QCP, as presented in this work.
The feasibility of coupling magnetic materials to cavity modes has been demonstrated in various cavity-magnonic systems~\cite{soykal2010strong,rameshti2022cavity, zuo2024cavity}.
Thus, there is good prospect for {potentially realizing our proposal on existing experimental platforms}~\cite{libersky2021direct, jarc2023cavity} 
with the externally applied magnetic field $h$ serving as a practical tuning parameter~\cite{kim2025observation}. 
Finally, also of interest in the present context are strange metals that exhibit strong quantum fluctuations~\cite{geiger2016investigations}.
{Our findings suggest that cavity coupling provides a means to access the elevated multipartite entanglement~\cite{Fang2025, Mazza2024} of such systems.}

\noindent {\bf{Outlook and summary.~~}} Our framework extends to driven-dissipative systems. 
In particular, cavity quantum systems have  emerged as attractive platforms for  realizing nonequilibrium phenomena where light plays a key role~\cite{schlawin2022cavity, hubener2021engineering, basov2025polaritonic, curtis2019cavity, ashida2020quantum,vinas2023controlling, jarc2023cavity,flores2025nonthermal}.
Both features discussed here---the suppression of $g_c$ and enhanced squeezing and entanglement---can be generalized to nonequilibrium settings, 
{which may also provide pathways for accessing the critically enhanced  squeezing and QFI identified in this work~\cite{dimer2007}}.
%%%
In closing the paragraph, we note that modeling a cavity as a single mode is an idealization, and realistic cavities {generally} require multi-mode modeling~\cite{roman2022effective}. 
Such a generalization is also necessary to systematically connect with the details of cavity design and the formal notion of a thermodynamic limit (see {Supplementary Note 4}). 
Generalizing to multi-mode settings, appropriate for cavity magnonics platforms, it can be shown that the  conclusions based on the single-mode model remains unchanged (see Methods and {Supplementary Note 4}).

To summarize,
we have theoretically demonstrated that quantum critical fluctuations in the matter sector greatly
amplify the response to the cavity-photon coupling and, especially,
promote the formation of a suprradiant state. 
This tendency is particularly striking when the cavity 
photons
directly couple to the
critical matter degree of freedom.
Here, in the quantum critical regime, the superradiant phase
becomes accessible 
{far below the ultrastrong coupling limit of cavity-matter interactions.}
{Moreover,} the system shows intrinsic squeezing 
and {enhanced} quantum Fisher information.
{In other words, coupling to the cavity photons provides a way to access
the elevated quantum entanglement of the underlying matter at quantum criticality, a finding that points a way towards realizing the potential of highly collective quantum materials for expanding the capacities of quantum information science.}
In this way, our work identifies a general  
principle for harnessing {matter} quantum criticality in cavity quantum materials 
 to realize  SRPTs at thermodynamic equilibrium  
 and
 {opens} a new route of investigation for 
 {designing} cavity {quantum} materials and  generating metrologically useful quantum states.

\medskip

%%%%%%%%%%%%%%
\noindent{\bf\large Methods}
\\
%%%%%%%%%%%%%%%%%%%%%%%%%%%%%%%%%%%%%%%%%%%%%%%%%%%%%%%%%%%%%%%%%%%%%%%%%%%%%%%%%%%%%%%%%%%%%%%%%%%%%%%%%%%%%%%%%%%%%%%%%%%%%%%%%%%%%%%%%%%
\noindent {\bf{Cavity-coupled TFIM -- additional properties:~~} }
For $\mathbf n \cdot \hat z = 0$,
$\hat H$ in 
Eqs.~(\ref{eq:fullH},\ref{eq:model}) has a $\mathbbm Z_2$ symmetry associated with $(a, \mathbf n \cdot \hat{\mathbf S}_{\mathbf r}) \to - (a, \mathbf n \cdot \hat{\mathbf S}_{\mathbf r})$~\cite{kirton2019introduction}.
(For $\mathbf n \cdot \hat z = 1$, the model lacks the $\mathbbm Z_2$-symmetry due to the applied magnetic field, and the ground state  supports a photon condensate at any non-vanishing model parameters.) 
In the $J=0$ limit, $\hat H$ reduces to the well-known Dicke model, and the system undergoes a spontaneous  $\mathbbm Z_2$-symmetry breaking as $g$ exceeds $g_c(J=0) = \sqrt{\omega_0 h}$.
The resultant superradiant phase  is characterized by a macroscopic occupation of the bosonic mode, $\expval{\hat a} \neq 0$, and a non-trivial spin-polarization, $\expval{\mathbf n \cdot {\mathbf{\hat S}}} \neq 0$.
In the opposite limit, $g=0$,  light and matter sectors are decoupled, and the ground state is the product state of the zero photon occupation state and the ground state of the TFIM.
Notably, the spin sector undergoes a ferromagnet to paramagnet (field-polarized state) quantum phase transition as the ratio $h/J$ is tuned across a critical value, $(h/J)_{\text{TFIM}}$. 

Here, we couple the cavity mode to the $\alpha$-th component of the net magnetization, $m_\alpha = \expval{\hat M_\pha}$ where $\hat M_\pha \equiv \sum_{\mathbf r} \hat S^\alpha_{\mathbf r}/N$ with $N$ being the total number of sites.
In the ferromagnetic (paramagnetic) phase $m_x \neq 0$ ($m_x = 0$).
At a fixed $J$, the static susceptibility, $\chi_x$,
diverges as $\chi_x \sim |h - h_\text{TFIM}|^{-\gamma}$ upon tuning $h$ across $h_\text{TFIM}$ with the critical exponent $\gamma$ being dimension-dependent.
This divergent susceptibility identifies $\hat M_x$ as the critical mode that is associated with the quantum phase transition in the TFIM.
The cavity mode directly couples to $\hat M_x$ for $\mathbf n = \hat x$.

\noindent{\bf The cavity coupling, gauge invariance and absence of no-go theorem:~~} The spins are 
Zeeman-coupled to a fluctuating magnetic field generated by the electromagnetic field inside the cavity,
\begin{align}
\hat H_{\hat a - \hat S} = \mu_B g_B \mathbf B(t) \cdot \sum_{\mathbf r} \hat{\mathbf S}_{\mathbf r} \, .
\end{align}
Upon quantization of the electromagnetic field,
the $\mathbf n$-th component of the total spin couples to the photon momentum quadrature $i(\hat a-\hat a^\dagger)$:
$\mu_B g_B \mathbf B(t) \to i g (\hat a - \hat a^\dagger) \mathbf n $, where $g = C \mu_B g_B $ is the effective light-matter coupling strength with $C$ being a parameter that depends on the details of the cavity~\cite{soykal2010strong,mckenzie2022theory, roman2022effective}.
After a canonical transformation $\hat a\to-i\hat a$ that swaps the photon position and momentum operators, Eq.~\eqref{eq:fullH} provides an equivalent description of the coupled system.

Unlike the electric-dipole interaction, which involves the vector potential $\mathbf A$ and requires a compensating diamagnetic term $(\propto \mathbf A\!\cdot\!\mathbf A)$ to preserve gauge invariance, the Zeeman coupling depends on the magnetic field $\mathbf B = \nabla\times\mathbf A$, which is itself gauge invariant under $\mathbf A \to \mathbf A + \nabla\chi$. Consequently, $\hat H_{\hat a - \hat S}$ 
is gauge invariant and no additional $A^2$-type term
operates.
This is in sharp contrast to the case of mobile two-level emitters or atoms with kinetic Hamiltonian $\hat H_{\text{spin}} = \hat{\mathbf p}^2/(2m)$~\cite{nataf2010no,roman2022effective}, where gauging via $\hat{\mathbf p}\to\hat{\mathbf p}-e\hat{\mathbf A}$ inevitably generates the diamagnetic term that leads to a no-go theorem against an  SRPT at equilibrium~\cite{rzazewski1975phase}
that continues to be discussed~\cite{kirton2019introduction}.
The manifestly gauge-invariant form of the Zeeman coupling means that our case is not subject to any no-go theorem.
A distinct miscroscopic perspective that reaches the same conclusion is presented in {Supplementary Note 6}.
 ~\par

\noindent {\bf{Cavity coupled to a critical degree of freedom:~~} }
We start from a large-$S$ analysis by introducing Holstein-Primakoff bosons with the  Ising paramagnetic state as the reference,  $\hat S^z_{\mathbf r} = S - \hat b_{\mathbf r}^\dagger \hat b_{\mathbf r}$ and $\hat S^-_{\mathbf r} = b_{\mathbf r}^\dagger \sqrt{2S - b_{\mathbf r}^\dagger \hat b_{\mathbf r}} $. 
Here, $\hat b_{\mathbf r}$ destroys the  quantum of spin-fluctuations transverse to the field-polarization direction---a ``magnon''---at site $\mathbf r$.
The effective Hamiltonian governing the resultant system of coupled photons and magnons is obtained from Eq.~\eqref{eq:fullH} by expanding about the large-$S$ saddle point and retaining terms up to order $S^0$, 
\begin{align}
\hat H_{\text{eff}} =& \omega_{0}\hat a^{\dagger}\hat a 
+ \sqrt{\frac{S}{2N}} g(\hat a +\hat a^{\dagger}) \sum_{\mathbf r} (b_{\mathbf r}+b_{\mathbf r}^{\dagger}) \nonumber\\
&   - \frac{S}{2} J \sum_{\expval{\mathbf r, \mathbf r'}} (\hat b_{\mathbf r}^{\dagger} \hat b_{\mathbf r'} + \hat b_{\mathbf r}\hat b_{\mathbf r'} + \mbox{h.c.}) 
+ {h} \sum_{\mathbf r}\hat b_{\mathbf r}^{\dagger} \hat b_{\mathbf r} \, .
\end{align}
We note two key features. 
First, the photons couple only to a {global} magnon operator.
This implies that, in the large-$S$ limit, only the $\mathbf k =0$ mode in the magnon sector is sensitive to the cavity coupling.
Second, as shown below, the magnon modes whose {BECs}  lead to the superradiant  and ferromagnetic phases, respectively, are in fact  {identical}.
This underscores the cooperation between the two phases. 

We isolate the $\mathbf k = 0$ magnon mode, and solve for the polaritonic normal modes supported by $\hat H_{\text{eff}}$.
We observe that the ferromagnetic exchange interaction, $J$, serves as an additional detuning parameter, 
such that the resonant regime is renormalized to $\omega_0 = h - z SJ$, where $z$ is the coordination number of the lattice on which $\hat H_{\text{spin}}$ is defined.
Using the Nambu basis $\hat \Phi = (\hat a \quad \hat{\mf b}_0 \quad a^\dagger \quad \hat{\mf b}_0^\dagger)$, where $\hat{\mf b}_{\mathbf k}$ is the $\mathbf k$-th Fourier mode of $\hat b_{\mathbf r}$, we find two branches in the Bogoliubov spectrum, 
\begin{align}
 E_{\pm}=\frac{1}{2\sqrt{2}}\sqrt{\Omega_+ \pm \sqrt{\Omega_-^2+8g^{2}hS\omega_{0}}}  \, , 
 \label{eq:E_PM}
\end{align}
where $\Omega_\pm = \omega_{0}^2 \pm  h (h - zSJ)$.
The vanishing of the dispersion of the `$-$' branch triggers a BEC in the corresponding polaritonic mode. 
The condition for the vanishing of $E_-$ determines the critical cavity-coupling for an SRPT,  
\begin{align}
g_c(h/J) = \sqrt{ \frac{\omega_0 h}{2S} \qty[1 - \frac{(h/J)_{\text{TFIM}}}{h/J} ]} \, ,
\label{eq:gc}
\end{align}
where $(h/J)_{\text{TFIM}} = z S$.
The phase boundary is shown in Fig.\ref{fig:ising0}{\bf a}.
We note that one would arrive at the same conclusion, by studying the pole structure of the dressed photon/cavity-mode propagator obtained by integrating our the collective spin mode, as discussed in Sec.~I.B of the SI.
At a fixed $J$, as $h \to h_{\text{TFIM}}^+$, $g_c$ vanishes with a mean-field exponent, $g_c \sim (h - h_{\text{TFIM}})^{1/2}$.
We note that $g_c = 0$ for $h \leq h_{\text{TFIM}}$ because the $\mathbbm{Z}_2$ symmetry of $\hat H$ is already broken by the ferromagnetic order.

We proceed to carry out DMRG calculations for the extreme quantum case, with $S=\frac{1}{2}$ and in one dimension. Details of the calculation are given in {Supplementary Note 2}.

\noindent {\bf{Quadrature-squeezing and quantum Fisher information:~~} }
We semi-classically determine the photon-magnon quadrature that is most squeezed in the vicinity of the SRPT. 
For this purpose, we derive (see {Supplementary Note 3})
%~\cite{sm}
an effective Hamiltonian that governs the excitations above a mean-field state specified by $\qty(\expval{\hat a}, \langle \hat{\mf b}_0 \rangle) = \sqrt{2 S N} (\alpha, -\beta)$ with $\alpha, \beta \geq 0$, 
\begin{align}
\delta \hat H_0' &= \omega_0 \delta \hat a^\dagger \delta \hat a 
+ h_{\text{eff}}(g, h, J)  \delta \hat{b}^\dagger \delta \hat{b}  + g_{\text{eff}}(g, h, J) (\delta \hat a + \delta \hat a^\dagger) (\delta \hat{b} + \delta \hat{b}^\dagger) \nn \\
&\quad + \Delta_{\text{pair}}(g, h, J) \qty(\delta \hat{b} \delta \hat{b}  +\delta \hat{b}^\dagger \delta \hat{b}^\dagger ) \, ,
\end{align}
where $\delta \hat a$ ($\delta \hat b$) is the fluctuation about $\expval{\hat a}$ ($\langle\hat {\mf b}_0 \rangle$) and the effective parameters are defined in Sec.III of the SI.
The quadrature in Eq.~\eqref{eq:X} is the general linear combination of $\delta \hat a$ and $\delta \hat b$ which results in a hermitian operator and it is analogous to a generalized position operator in simple harmonic oscillators~\cite{hayashida2023perfect} (see {Supplementary Note 3}).

We use the {Optim.jl} package in {Julia} to numerically determine the set of angles $\left. (\theta, \psi, \phi)\right|_{\text{min}}$ for which the variance of $\hat X_{ \mathrm{\theta, \psi, \phi }}$ is lowest.
We refer to this operator as $\hat X_{\text{min}}$.

Because of Heisenberg's uncertainly relations, there must exist an operator $\hat X_{\text{max}}$ that is conjugate to $\hat X_{\text{min}}$: $\qty[\hat X_{\text{min}}, \hat X_{\text{max}}] = -\frac{i}{2}$, and whose variance is maximized. 
The same computational method leads us to the needed set of angles $\left. (\theta, \psi, \phi)\right|_{\text{max}}$ for which the variance of $\hat X_{ \theta, \psi, \phi }$ is the highest.
This operator is the desired $\hat X_{\text{max}}$.
It can {be} checked that the product of the two variances equals $1/16$, satisfying the lower bound of the uncertainly relation for variances of bosonic mode operators~\cite{meystre2007elements}. 

{In principle, the} large variance of $\hat X_{\text{max}}$ can be utilized as a resource for high-precision parameter estimation~\cite{fadel2024quantum}.
In particular, we consider unitarily imprinting the parameter $\vartheta$ as  $\ket{\psi_0} \to \ket{\psi(\vartheta)} = \exp{-i \vartheta \hat X_{\text{max}}} \ket{\psi_0}$. 
The Cram\'{e}rs-Rao bound dictates that for $m$ independent measurements of $\vartheta$, its variance $\Delta \vartheta^2 \geq 1/(m F_Q(\hat X_{\text{max}}))$, where $F_Q(\hat X)$ is the quantum Fisher information associated with the operator {$\hat X_{\text{max}}$} in the state $\ket{\psi_0}$~\cite{braunstein1994statistical}.
Assuming $\ket{\psi_0}$ is a pure state, one obtains $F_Q(\hat X_{\text{max}}) = 4 \Delta \hat X_{\text{max}}^2$ (see {Supplementary Note 3}), which implies, in principle,  $\Delta \vartheta^2$ can be reduced to zero.

\noindent {\bf{Cavity coupling to a non-critical mode.~~}}
As described in the main text, we also consider the case of 
cavity coupling to a non-critical model, corresponding to $\mathbf n = \hat y$. From the 
large-$S$ limit, we can see that 
the cavity mode directly couples to $\sum_{\mathbf r} (b_{\mathbf r} - b_{\mathbf r}^{\dagger})$.
Therefore,  the $\mathbf k= 0$ magnon mode that must condense to produce a superradiant phase, $(\hat{\bar b}_{\mathbf 0} - \hat{\bar b}_{\mathbf 0}^\dagger)/i$, is distinct
from the ferromagnetic order parameter of the pure
matter sector, $ (\hat{\bar b}_{\mathbf 0} + \hat{\bar b}_{\mathbf 0}^\dagger)$.
Because 
these two spin modes are orthogonal, 
phase transitions in the two sectors remain decoupled at the leading order in the large-$S$ limit. 

Instead of pursuing higher order corrections in $1/S$, here, we focus on $d=1$ with $S=1/2$ and derive an analytically exact free energy in terms of $\expval{a}$~\cite{lee2004,gammelmark2011phase}.
In this approach, $\expval{a}$ is treated as a real-valued order parameter, $\langle a\rangle \equiv \sqrt{N}\phi/2$, and the spin degrees of freedom are integrated out to obtain the ground state energy density, $\mc E_{g}(\phi)$ (see {Supplementary Note 4}). 
Minimizing $\mc E_{g}$ with respect to
$\phi$ yields the phase diagram shown in Fig.~\ref{fig:ising}{\bf a}. 

For $h/J < 1/2$, the superradiant transition is discontinuous; at the TFIM critical point, $(h/J)_{\text{TFIM}}=1/2$, the Ising order vanishes while the superradiant transition remains discontinuous with a reduced critical coupling  $g_{\mathrm{c}} \approx 0.87 \sqrt{J\omega_{0}}$ (vs. $g_{\mathrm{c}}\approx 0.92\sqrt{J\omega_{0}}$ at $h=0$). 
For $h/J > 1/2$, the ground-state energy expands as $\mc E_g/J = r \phi^2 + u \phi^4 + v \phi^6$, with $r=(g_{\mathrm{c}}^{2}-g^{2})/4J^{2}$ and $g_{\mathrm{c}}=\sqrt{J\omega_{0}/f(h/J)}$, where the coefficients $f,u,v$ are shown in Sec IV of the SI.
In the range $(h/J)_\text{TFIM} < h/J < (h/J)_{\text{tri}} \approx 0.55$, we find $u<0$, $v>0$, yielding a discontinuous SRPT that extends the first-order transition line from $h/J \le (h/J)_\text{TFIM}$. 
For $h/J > (h/J)_{\text{tri}}$, $u>0$ and the SRPT becomes continuous, smoothly connecting to the Dicke limit ($h/J, g/\sqrt{\omega_0 J}\to \infty$ at fixed $g/\sqrt{\omega_0 h}\sim 1$). 
The intersection between the two types of QPTs at $h/J = (h/J)_{\text{tri}}$ defines the tricritical point, whose scaling we analyze below. 

In the vicinity of the tricritical point with a fixed $J$ and $\omega_0$,  $u \propto (h - h_{\text{tri}})/J$ and the number of photons in the condensate obtains the scaling form 
\begin{align}
\mc N = \phi^2 = \qty(g/g_c - 1)^{\frac{1}{2}} f_{\mc N}\qty(\frac{u/\sqrt{v}}{\sqrt{g/g_c - 1} } ) \, .
\label{eq:N}
\end{align} 
Here, the dimensionless function $f_{\mc N}(x) =c_1/\qty[c_2 x + \sqrt{1 + (c_2 x)^2}]$ with $c_n$'s being dimensionless parameters, and it  has the limiting behaviors, $\lim_{x \to 0} f_\mc{N}(x) \sim 1$ and $\lim_{x \to \infty} f_\mc{N}(x) \sim 1/x$.
Therefore, in the superradiant phase at $h = h_\text{tri}$, $\mc N \sim (g/g_c - 1)^\half$  , while $\mc N \sim (g/g_c - 1)$ for $h > h_\text{tri}$.
While the latter is the standard mean-field result, the former is a peculiarity of tricritical points which was also 
observed in a variant of the pure Dicke model~\cite{Xu2019}.
As shown in Fig.~\ref{fig:ising}{\bf b}, $f_\mc{N}$ controls the crossover between the two scaling limits with the crossover scale determined by the condition $c_2 x = 1$.

%%%%%%%%%%%%%%%%%%%%%%%%%%%%%
%%%%%%%%%%%%%%%%%%%%%%%%%%%%%

\noindent {\bf{Phase diagram of the XY spin model: ~~}}
The ferromagnetic XY model realizes a ferromagnetic phase with a net magnetization along $\hat x$ ($\hat y$) for $\Delta > 0$ ($\Delta < 0$), which spontaneously breaks the $\mathbbm Z_2$ symmetry of $\hat H_\text{spin}$, present at any $\Delta \neq 0$~\cite{Lieb1961}.
The QPT between the two phases is continuous with the QCP at $\Delta = 0$ realizing an enhanced $SO(2)$ symmetry. 
Unitarily equivalent Ising models are recovered in the limits $\Delta \to \pm 1$.

\noindent {\bf{Thermodynamic limit and multi-mode generalization: ~~}} The quantum phase transitions discussed in this work implicitly assume a thermodynamic limit in which both the quantum magnet and the cavity participates~\cite{roman2022effective}, as discussed in {Supplementary Note 4}. 
If strictly implemented, this will result in a quasi-continuum of photonic modes which is expected to limit the regime of validity of the single-mode approximation, especially if the matter sector is in a charge-itinerant phase~\cite{Eckhardt2025}.
While we formally focus on the strong coupling limit of the cavity coupled Hubbard model~\cite{sentef2020quantum, mazza2019superradiant, passetti2023},  where charge fluctuations are localized and the matter sector is a quantum magnet (see {Supplementary Note 4}), multi-mode aspects still require careful considerations. 
In {Supplementary Note 4}, we discuss the limits of the single-mode approximation and demonstrate that the multi-mode version of our model, as applicable to cavity magnonics settings~\cite{rameshti2022cavity}, continue to exhibit qualitatively similar physics.
We summarize the results below.

The simplest multi-mode generalization of the cavity coupled to a critical degree of freedom is governed by the Hamiltonian
\begin{align}
H_\mathrm{multi} = \sum_{k=1}^M \omega_k a_k^\dagger a_k - h S^z_T + \frac{1}{\sqrt{N}} \sum_{k=1}^M g_k (a_k + a_k^\dagger) S^x_T + H_\mathrm{spin},
\label{eq:multimode}
\end{align}
where $k$ labels the $M$ modes present within the cavity, $\omega_k$ specifies the mode profile, $g_k$ is the distribution of collective light-matter coupling for the $k$-th mode, and $S^\mu_T = \sum_j S^\mu_j$ is the $\mu$-th component of the collective spin. 
For our minimal requirement of the existence of a quantum phase transition in the spin sector, the spin-spin interaction part of the Hamiltonian is assumed to have a collective form,  $H_\mathrm{spin} = -(J/N) (S^x_T)^2$.
As detailed in Sec.~VI of the SI, the condition for SRPT in this multi-mode models is $\sum_{k=1}^M g_k^2/\omega_k = (h-J)$, which implies that with growing proximity to the magnetic QCP it becomes easier to obtain a superradiant state. 
This is in qualitative agreement with our single-mode model. 

%%%%%%%%%%%%%%%%%%%%%%%%%%%%%%%%%%%%%%%%%%%%%%%%%%%%%%%%%%%%%%%%%%%%%%%%%%%%%%%%%%%%%%%%%%%%%%%%%%%%%%%%%%%%%%%%%%%%%%%%%%%%%%%%%%%%%%%%%%%%%%%%%%%%%%

\vskip 0.5 cm
\noindent{\bf\large Data availability}
\\
{
The data that support the findings of this study are either presented in the manuscript or available at }\href{https://doi.org/10.5281/zenodo.19655430}{{https://doi.org/10.5281/zenodo.19655430}}.

\vskip 0.5 cm
\noindent{\bf\large Code availability}
\\
The computer codes that were used to generate the data that support the findings of this study are available from the corresponding author upon request.

%%%%%%%%%%

\bibliographystyle{naturemagallauthors}
%%%%%%%%%%%%%%%

\bibliography{cavity_sorted.bib}

\clearpage

\noindent{\bf Acknowledgements}
\\
We thank Songtao Chen, Kaden Hazzard, Jun Kono, 
B. Prasanna Venkatesh, Han Pu, Vaibhav Sharma, and Hanyu Zhu for useful discussions. 
Work at Rice has primarily been supported by the 
DOE, BES Grant No.\ DE-SC0026179, the AFOSR Grant No.\ FA9550-21-1-0356,
the Robert A. Welch Foundation Grant No.\ C-1411, 
and 
the Vannevar Bush Faculty Fellowship ONR-VB N00014-23-1-2870. 
The
majority of the computational calculations have been performed on the Shared University Grid
at Rice funded by NSF under Grant EIA-0216467, a partnership between Rice University, Sun
Microsystems, and Sigma Solutions, Inc., the Big-Data Private-Cloud Research Cyberinfrastructure
MRI-award funded by NSF under Grant No. CNS-1338099, and the Extreme Science and
Engineering Discovery Environment (XSEDE) by NSF under Grant No. DMR170109. 
S.P. acknowledges funding by the European Union (ERC Adv. Grant CorMeTop, project 101055088), the Austrian Science Fund (FWF) through
the projects SFB F 86 (Q-M\&S), FOR 5249 (QUAST), and 10.55776/COE1 (quantA), and the US AFOSR through project FA8655-24-1-7018 (CorTopS).
Q.S. acknowledges the hospitality of the Aspen Center for Physics, 
which is supported by NSF grant No. PHY-2210452.

\vspace{0.2cm}
\noindent{\bf Author contributions}
\\
Q.S. conceived the research. S.S., Y.W., M.M. and Q.S. carried out model studies.
S.S., Y.W. and Q.S. wrote the manuscript, with inputs from all 
{the other} authors 
{(M.M. and S.P.)}.
%\\

\vspace{0.2cm}
\noindent{\bf Competing 
 interests}\\
The authors declare no competing 
 interests.
 %\\
 
 \vspace{0.2cm}
 \noindent{\bf Additional information}\\
Correspondence and requests for materials should be addressed to 
Q.S. (qmsi@rice.edu).

\clearpage

\vskip 0.1 in
\noindent
{\large \bf{Figure Legends/Captions } }

\clearpage

\begin{figure}[h!]
\centering
\includegraphics[width=1.0\columnwidth]{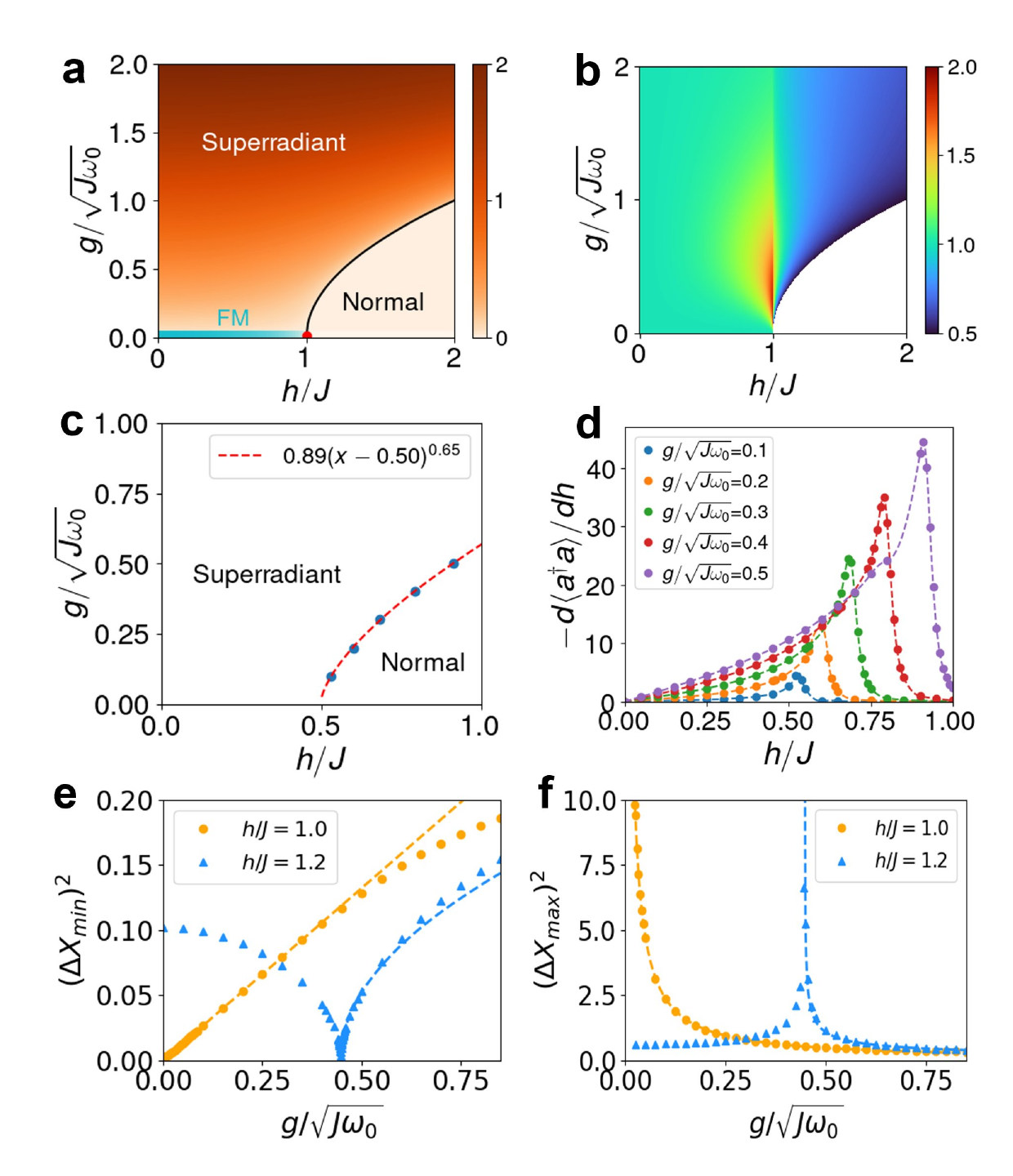}
\caption{{\bf
Phase diagram, squeezing and entanglement for a cavity mode  directly coupled to quantum critical mode. ~~~}
{\bf a}, Phase diagram in the large-$S$ limit, depicting $\expval{\hat a}$; black curve denotes continuous {SRPTs} that terminates at the TFIM QCP (red dot); blue bar refers to the ferromagnetic (FM) order. We set the lattice coordination number  $z=2$ and 
%,\,
spin $S=1/2$.
}
%%%%%%%%%%%%%%%%%%%%%%%%%%%%%%%%%%%%%%%%%%%%%%%
%%%%%%%%%%%%%%%%%%%%%%%%%%%%%%%%%%%%%%%%%%%%%%%
\end{figure}
\begin{figure}[t]
  \contcaption{
 (cont'd) 
 {\bf b}, Scaling of $\expval{\hat a}$ with $(g - g_c)^{\beta}$ in the large-$S$ limit (color bar represents $\beta$): $g_c = 0$ for $h/J \leq (h/J)_\text{TFIM}$; the Dicke model gives $\beta = 1/2$. 
{\bf c}, Phase boundary from DMRG simulations (blue points) with fit $g_c \propto (h-h_\text{TFIM})^{0.65}$.
{\bf d}, Data in (c) is extracted from the peaks of $\partial \expval{\hat a^\dagger \hat a}/\partial h$ as a function of $h$. 
{We observe a reduced participation of the cavity mode in the $h$-tuned quantum phase transition, indicating a smooth evolution of the SRPT to the conventional TFIM quantum phase transition as $g\to 0$.}
{\bf e},{\bf f},  Minimum and maximum variances of the quadrature 
$\hat X_{\mathrm{\theta, \psi, \phi}}$ (cf.~Eq.~\eqref{eq:X}) along the phase boundary in {\bf a}. 
The minimum variance $(\Delta X_\text{min})^2$ vanishes, indicating 
perfect intrinsic squeezing, while the maximum variance 
$(\Delta X_\text{max})^2$ diverges, indicating 
enhanced quantum entanglement. Dashed lines show fits to numerical data as 
$g \to g_c^+$: at $h=J$ ($h>J$), $(\Delta X_\text{min})^2 \sim g$ 
($\sqrt{g-g_c}$), and $(\Delta X_\text{max})^2 \sim g^{-1}$ 
($(g-g_c)^{-1/2}$).
Note that $(h/J)_\text{TFIM}$ in the large-$S$ limit is distinct than the fully quantum solution {[cf. {\bf a} vs. {\bf c}]}; 
we have set $\omega_0 = 1 = J$ in {\bf c} -- {\bf e}.
}
\label{fig:ising0}
\end{figure}

\clearpage

\begin{figure}[h!]
\centering
\includegraphics[width=.75\columnwidth]{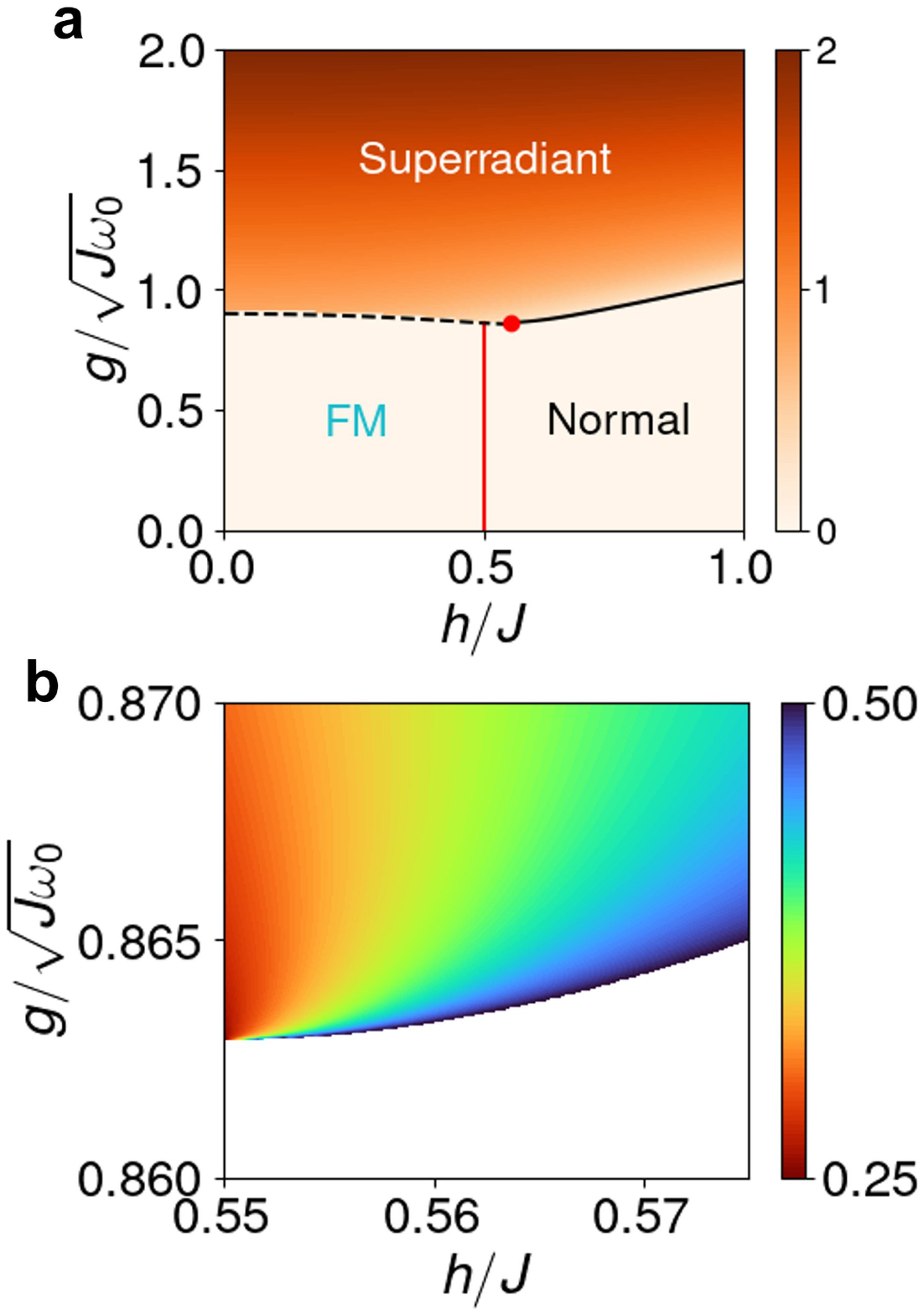}
\caption{{\bf Phase diagram for a cavity mode coupled 
to non-critical degrees of freedom.~~~}
{\bf a}, Analytically obtained exact phase diagram {in the full quantum limit of $S = 1/2$} where the color bar indicates $\expval{\hat a}$; red solid line marks the {QCP} of the one-dimensional {TFIM};  dashed (solid) black lines denote discontinuous (continuous)  {SRPTs}; red dot represents the tricritical point. 
{\bf b},  Scaling of $\expval{\hat a}$ with $(g - g_{\mathrm{c}})^{\beta}$ in the vicinity of the tricritical point at $h/J \approx 0.55 $. 
The crossover behavior is dictated by Eq.~\eqref{eq:N}. 
}
\label{fig:ising}
\end{figure}

\clearpage

\begin{center}
\textbf{\large Supplementary Information for: 
Amplified response of cavity-coupled quantum-critical systems}
\\%[0.6cm]
[0.3cm]

Shouvik\ Sur$^{1,\dagger}$,
Yiming\ Wang$^{1,\dagger}$,
Mounica\ Mahankali$^{1}$,
Silke\ Paschen$^{2,1}$,
Qimiao\ Si$^{1,\ast}$ 
\\[0.3cm]

$^1$Department of Physics and Astronomy, Extreme Quantum Materials Alliance, Smalley-Curl Institute, Rice University, Houston, Texas 77005, USA\\[-0.cm]

$^2$Institute of Solid State Physics, Vienna University of Technology, Wiedner Hauptstr. 8-10, 1040
Vienna, Austria

\normalsize{$^\ast$To whom correspondence should be addressed; E-mail:  qmsi@rice.edu.}

\end{center}
\setcounter{secnumdepth}{2} % default value for 'report' class is "2"
\setcounter{equation}{0}
\setcounter{figure}{0}
\setcounter{table}{0}
\renewcommand{\theequation}{S\arabic{equation}}
\renewcommand{\thefigure}{S\arabic{figure}}
\renewcommand{\thesection}{Supplementary Note~\arabic{section}:}

\clearpage
%\onecolumngrid
\setcounter{secnumdepth}{3}

{
\section{\label{app:crit-mode} Analytical results for cavity coupled to  the critical mode of TFIM}
In this section we provide the details of analytical calculations that inform our results on the cavity mode coupled to a critical mode in the {TFIM}.

\subsection*{\label{app:largeS}  Large-$S$ treatment of cavity coupled to  the critical mode of TFIM}
}
In this section we provide the details of the large-$S$ based mean-field calculation that led to Figs. 2a and 2b in the main section.
The Hamiltonian is given by   
\begin{align}
    H = -J\sum_{\langle i,j\rangle} S_{i}^{x} S_{j}^{x} - h \sum_{i} S_{i}^{z} 
    + \frac{g}{\sqrt{N}} (a + a^{\dagger}) \sum_{i} S^{x}_{i} + \omega_{0} a^{\dagger}a.
\end{align}
Here $N$ is the number of spin sites; the normalization in $g/\sqrt{N}$ ensures that the system has a well-defined thermodynamic limit~\cite{emary2003chaos}. 
In the large-$S$ limit, the spin-$\frac{1}{2}$ operators are generalized into spin-$S$ operators, allowing for a controlled $1/S$ expansion. The quantum fluctuations are suppressed as 
$1/S$, and the spin dynamics become effectively classical. The mean-field approximation thus corresponds to the leading-order (classical) term of the large-$S$ expansion.

We decouple
the Ising interaction,  
\begin{align}
    \sum_{\langle i,j \rangle} S_{i}^{x} S_{j}^{x} 
    \;\approx\; z m_{x} \sum_{i} S_{i}^{x} - \frac{zN}{2} m_{x}^{2} \, ,
\end{align}
where $m_{x} = \langle S_{i}^{x} \rangle$ denotes the uniform magnetization and $z$ is the coordination number. For square lattie in $d$-dimensions, $z=2d$.  

By introducing a coherent-state representation for the photons in the partition function, we obtain  
\begin{align}
    Z = \int D[a^{\dagger},a]\, e^{-\beta \omega_{0} a^{\dagger} a} \,
    \mathrm{Tr}_{\text{spin}} \, e^{-\beta H_{a}} \, ,
    \label{partition}
\end{align}
with the effective spin Hamiltonian  
\begin{align}
    H_{a} = -h \sum_{i} S_{i}^{z} 
    - \biggl[zJ m_{x} - \frac{g}{\sqrt{N}} (a + a^{\dagger}) \biggr] \sum_{i} S_{i}^{x} \, .
\end{align}

The resulting free-energy density is  
\begin{align}
    f = -\frac{1}{\beta N} \ln Z
      = \frac{1}{4}\,\omega_{0}\phi^{2} + \frac{1}{2} zJ m_{x}^{2} 
        - \sqrt{\,h^{2} + (g\phi - zJ m_{x})^{2}\,} \, ,
\end{align}
where $\phi = \langle a + a^{\dagger} \rangle / \sqrt{N}$ is the superradiant order parameter.  

Minimization of $f$ with respect to $\phi$ and $m_{x}$ yields the saddle-point equations,  
\begin{align}
    \tfrac{1}{2}\omega_{0}\phi &= gS \, \frac{g\phi - zJ m_{x}}{\sqrt{\,h^{2} + (g\phi - zJ m_{x})^{2}}} \, , \\
    m_{x} &= -S \, \frac{g\phi - zJ m_{x}}{\sqrt{\,h^{2} + (g\phi - zJ m_{x})^{2}}} \, .
\end{align}

The solutions take the form  
\begin{align}
    \phi &= 
    \begin{cases}
        \dfrac{2g}{\omega_{0}} \, 
        \dfrac{\sqrt{S^{2}\bigl(2g^{2}/\omega_{0}+zJ\bigr)^{2} - h^{2}}}{\,2g^{2}/\omega_{0}+zJ} \, , 
        & h < S\bigl(2g^{2}/\omega_{0}+zJ\bigr)\, , \\[1.2em]
        0 \, , & \text{else} \, ,
    \end{cases} \\[0.6em]
    m_{x} &= 
    \begin{cases}
        - \dfrac{\sqrt{S^{2}\bigl(2g^{2}/\omega_{0}+zJ\bigr)^{2} - h^{2}}}{\,2g^{2}/\omega_{0}+zJ} \, , 
        & h < S\bigl(2g^{2}/\omega_{0}+zJ\bigr) \, ,\\[1.2em]
        0 \, , & \text{else} \, .       
    \end{cases}
\end{align}

The superradiant order parameter $\phi$ is shown in Fig.~2{\bf a}, and its scaling with the coupling $g$ is presented in Fig.~2{\bf b}.  
\\

{
\subsection*{The cavity-mode's perspective of the superradiant quantum phase transition\label{app:GF}}
Spin fluctuations in the quantum magnet dresses the dynamics of the cavity mode.
We capture this physics by determining the effective propagator of $\hat a(t)$ by integrating out the spin fluctuations. 
In order to concretely demonstrate how the spin susceptibility close to the magnetic QCP affects the softening condition in the cavity photon's dynamics, it is sufficient to consider the Lipkin-Meshkov-Glick (LMG) model coupled to a single cavity mode,
\begin{align}
H = \omega_0 a^\dagger a + \frac{g}{\sqrt{N}} (a + a^\dagger) S^x_T - \frac{J}{N} (S^x_T)^2 - h S^z_T,
\end{align}
where $S^\mu_T = \sum_{j=1}^N S^\mu_j$ is the $\mu$-th component of the total spin, and the last two terms represents the LMG model.
The Holstein-Primakoff (HP) transformation of $S^\mu_T$ is asymptotically exact in the $N\to \infty$ limit and the dynamics of $H$ is accurately captured by a Gaussian theory of the cavity mode and the HP bosons.
The action is defined as
\begin{align}
    S=\int dt a^{\dagger}(i\partial_{t}-\omega)a+b^{\dagger}(i\partial_t-h)b+\frac{J}{4}(b+b^\dagger)^2+\frac{g}{2}(a+a^\dagger)(b+b^\dagger).
\end{align}
Integrating out the HP bosons leads to the effective time-ordered propagator $D_{x}^{a}(t)\equiv-i\langle \mathcal{T}x_{a}(t)x_{a}(0)\rangle$ for the $x_a = (a + a^\dagger)/\sqrt{2}$ quadrature of the cavity mode, 
\begin{align}
D_x^{(a)}(\omega) = \frac{\omega_0}{(\omega^2-\omega_0^2)-\frac{1}{2}g^2\omega_0\chi(\omega)}
\end{align}
where $\chi(\omega)\equiv -\frac{i}{N}\int dt e^{i\omega t}\langle \mathcal{T}S^{x}_{T}(t)S_{T}^{x}(0)\rangle = \frac{2h}{\omega^2-h(h-J)}$ is the collective spin susceptibility for the LMG model, and $\mathcal{T}$ is the time-ordered operator. 
We note that the existence of a quantum phase transition in the LMG model is captured by the divergence of $\chi(\omega \to 0)$ at $h = J$.
Further, $D_x^{(a)}(\omega)$ vanishes at the poles of $\chi(\omega)$, i.e. the zeros of $D_x^{(a)}(\omega)$ are the collective spin excitations in the system. 

The  poles of $D(\omega)$ represent  spin-fluctuation-dressed photons, i.e. polaritons, and they are located at 
\begin{align}
\omega = \omega_\pm = \frac{1}{\sqrt{2}}\sqrt{
 h(h - J)+\omega_0^2 \pm \sqrt{[h(h - J)-\omega_0^2]^2 + 4 g^2h\omega_0}.
}
\label{eq:dLMG}
\end{align}
The vanishing of the $\omega_-$ branch at $g = g_c \equiv \sqrt{\omega_0(h-J)}$ signals the onset of condensation into the corresponding polaritonic mode  and the concomitant SRPT.
We note that in defining  $g_c$ we have assumed that $h$ and $\omega_0$ are fixed and $g$ is tuned to drive the SRPT (i.e. traversing vertically in the phase diagram in Fig.~2a of the manuscript). 
In order to reflect the most likely experimental design for achieving such SRPT, we can equivalently fix $g$ and tune $h$ to achieve the SRPT (i.e., traversing horizontally in the phase diagram in Fig.~2a of the manuscript), which will lead to a critical magnetic field, $h_c = J + g^2/\omega_0$, associated with the vanishing of $\omega_-$.

\section{DMRG calculations}
In this section we provide key technical details on the DMRG calculations. 
Calculations were performed using the TeNPy Library (version 1.0.5)\cite{tenpy2018}. Using TeNPy's \texttt{IrregularLattice} class, the system is constructed by manually adding a single bosonic site to a spin-1/2 chain. We considered a 100 site chain with a bosonic site with a cap of 100 on it's occupation number. Given a set of Hamiltonian parameters, the ground state is calculated using TeNPy's \texttt{TwoSiteDMRGEngine} with a random product state as the initial trial wavefunction. For all the DMRG runs, the bond dimension has been increased in increments of 50 with each sweep until a maximum of 2000 is reached.

Fig.\ref{fig:h-N-DickeTFIM} is generated by collecting ground states at each point in the parameter space and calculating relevant operator averages. 

\subsection*{Determination of the phase boundary}

Fig.\ref{fig:h-N-DickeTFIM}{\bf a} shows the boson occupation number as a function of $h$ for certain values of $g$. Assuming that from the superradiant side $\left\langle  a^{\dag}a \right\rangle \sim (h_c \left( g \right) - h)^{\alpha}$ as $h\rightarrow h_c \left( g \right)$ with $\alpha < 1$, the derivative $-\frac{d\left\langle  a^{\dag}a \right\rangle}{dh}$ peaks at $h=h_{c}\left( g \right)$. Using this observation, we calculated $h_c\left(g\right)$ for every $g$ by using a cubit spline interpolation for $\left\langle a^{\dag}a \right\rangle$ for every $g$ in figure \ref{fig:h-N-DickeTFIM}{\bf a} and then calculating it's derivative. Figure \ref{fig:h-N-DickeTFIM}{\bf b} demonstrates this for $g=0.3\sqrt{J\omega_0}$. The extracted $h_c \left(g\right)$ for each $g$ is plotted in figure 2{\bf c} whose fitting to curve $C(x-x_0)^{\alpha}$ gave the parameters  $C=0.887256$, $x_0=0.495780$, $\alpha=0.64967$.  

We close by noting that the location of the TFIM QCP in $d=1$, the fully quantum limit, is half of that found in
large-$S$ analyses.

\begin{figure*} \centering
\includegraphics[width=\textwidth]{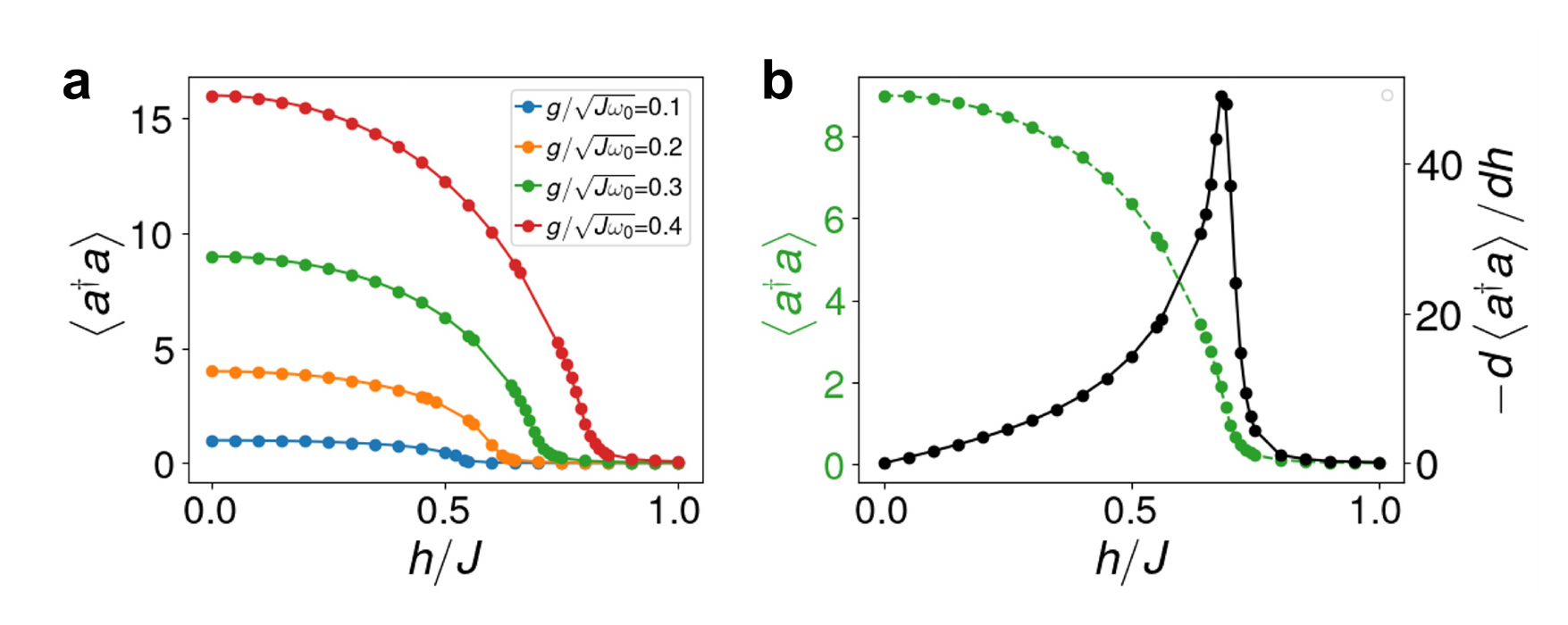}
\caption{{\bf DMRG results of transverse field Ising chain of 100 sites coupled to cavity photons with $J=1$, $\omega_0=1$. ~~~} {\bf a}, $\left \langle a^{\dag}a \right \rangle$ as a function of $h$. {\bf b}, $\left \langle a^{\dag}a \right \rangle$ with it's cubic spline interpolation (dashed line) and it's first derivative with respect to $h$ at $g=0.3\sqrt{J\omega_0}$ showing that it peaks near $h_c$.
\label{fig:h-N-DickeTFIM}}
\end{figure*}

\section{Intrinsic squeezing and QFI in the large-$S$ limit}
In this section we provide the details on the computation of the squeezed quadrature and additional details on quantum Fisher information for the case of the cavity mode coupled to the critical mode of the TFIM. 
The Hamiltonian is given by  
\begin{align}
\hat H = \omega_0 \hat a^\dagger \hat a 
- h\sum_{j=1}^N \hat S_j^z 
+ \frac{g'}{\sqrt{2SN}} (\hat a + \hat a) \sum_{j=1}^N \hat S_j^x
- \frac{J'}{2S} \sum_{j=1}^{N-1} \hat S_j^x \hat S_{j+1}^x  \, .
\label{eq:ham-long}
\end{align}
where we have introduced the scaled couplings  $$g'=g\sqrt{2S},\qquad J'=2JS$$ for notational convenience. Introducing Holstein-Primakoff (HP) bosons,
\begin{align}
\hat S_j^z = S - \hat n_j \, ; \qquad
\hat S_j^+ = \sqrt{2S - \hat n_j} ~\hat b_j \, ; \qquad 
\hat S_j^- = \hat b_j^\dagger ~\sqrt{2S - \hat n_j} \, ,
\end{align}
we obtain
\begin{align}
 \hat H + h N S =&~ \omega_0 \hat a^\dagger \hat a 
+ h \sum_{j=1}^N \hat n_j
+ \frac{g'}{2\sqrt{2SN}} (\hat a + \hat a) \sum_{j=1}^N (\sqrt{2S - \hat n_j} \hat b_j + \hat b_j^\dagger \sqrt{2S - \hat n_j})  \nn \\
&~ - \frac{J'}{4(2S)} \sum_{j=1}^{N-1}  \hat  (\sqrt{2S - \hat n_j} \hat b_j + \hat b_j^\dagger \sqrt{2S - \hat n_j})  (\sqrt{2S - \hat n_{j+1}} \hat b_{j+1} + \hat b_{j+1}^\dagger \sqrt{2S - \hat n_{j+1}})  \\
\equiv &~ H' \, .
\end{align}
(Note that the Fock space at site $j$ is $(2S+1)$-dimensional, supporting a maximum of $2S$ HP bosons. The count reduces to that for hardcore bosons as $S \to 1/2$.)
We will assume that both $\expval{\hat n_j}$ and $\delta \hat n_j$ are sufficiently weaker than $2S$ such that 
\begin{align}
 \sqrt{2S - \hat n_j}  \to \sqrt{2S} - \frac{\hat n_j}{2 \sqrt{2S}} + \order{\sqrt{2S} \qty(\frac{\hat n_j}{2S})^2} \, .
\end{align}
Therefore,
\begin{align}
& \hat H' = \omega_0 \hat a^\dagger \hat a 
 + h \sum_{j=1}^N \hat n_j
+ \frac{g'}{2\sqrt{N}} (\hat a + \hat a) \sum_{j=1}^N (\hat b_j + \hat b_j^\dagger) 
- \frac{g'}{(2\sqrt{2S})^2\sqrt{N}} (\hat a + \hat a) \sum_{j=1}^N (\hat n_j \hat b_j + \hat b_j^\dagger \hat n_j) \nn \\
&\quad ~ - \frac{J'}{4} \sum_{j=1}^{N-1}  (\hat b_j + \hat b_j^\dagger) (\hat b_{j+1} + \hat b_{j+1}^\dagger) \nn \\
&\quad ~ + \frac{J'}{8(2S)} \sum_{j=1}^{N-1}  \qty[
(\hat b_j + \hat b_j^\dagger)(\hat n_{j+1} \hat b_{j+1} + \hat b_{j+1}^\dagger \hat n_{j+1}) 
+ (\hat n_{j} \hat b_{j} + \hat b_{j}^\dagger \hat n_{j}) (\hat b_{j+1} + \hat b_{j+1}^\dagger) 
] + \ldots
\end{align}
where the ellipses represent higher order terms that are irrelevant for our analysis.

Next, we Fourier transform the HP bosons,
\begin{align}
\hat b_j = \frac{1}{\sqrt{N}} \sum_k e^{i k j} \hat{\mf b}_k \qquad \& \qquad
\hat{\mf b}_k = \frac{1}{\sqrt{N}} \sum_j e^{-i k j} \hat b_j  \qquad \mbox{with} \qquad
\delta_{k, k'} \equiv \frac{1}{N} \sum_j e^{i (k-k')j}
\end{align}
to obtain
\begin{align}
\hat H' =&~ \omega_0 \hat a^\dagger \hat a 
 + h \sum_{k} \hat{\mf n}_k
+ \frac{g'}{2} (\hat a + \hat a) (\hat{\mf b}_0 + \hat{\mf b}_0^\dagger) 
- \frac{g'}{(2\sqrt{2SN})^2} (\hat a + \hat a) \sum_{k,p} \qty(\hat{\mf b}_{k+p}^\dagger \hat{\mf b}_k  \hat{\mf b}_p 
+ \mbox{h.c.}
)  \nn \\
&~ - \frac{J'}{4} \sum_{k}   \qty( e^{-ik} \hat{\mf b}_k \hat{\mf b}_k^\dagger + e^{ik} \hat{\mf b}_{-k} \hat{\mf b}_{k} + \mbox{h.c}) \nn \\
&~ + \frac{J'}{8(2SN)} \sum_{k,q,p}  \qty[
e^{-ik} \qty(
\hat{\mf b}_k  \hat{\mf b}_{k+q+p}^\dagger \hat{\mf b}_q \hat{\mf b}_p 
+ \hat{\mf b}_k \hat{\mf b}_q^\dagger \hat{\mf b}_{k+p-q}^\dagger \hat{\mf b}_p  
)
+ e^{ik} \qty( 
\hat{\mf b}_k^\dagger \hat{\mf b}_{p+q-k}^\dagger \hat{\mf b}_q  \hat{\mf b}_p 
+ \hat{\mf b}_k^\dagger \hat{\mf b}_q^\dagger \hat{\mf b}_{p-k-q}^\dagger \hat{\mf b}_p
)
] \nn \\
&~ + \frac{J'}{8(2SN)} \sum_{k,q,p}  \qty[
e^{iq} \qty(
 \hat{\mf b}_k^\dagger \hat{\mf b}_q^\dagger \hat{\mf b}_p \hat{\mf b}_{p-k-q}^\dagger 
+ \hat{\mf b}_k^\dagger \hat{\mf b}_q  \hat{\mf b}_p \hat{\mf b}_{p+q-k}^\dagger 
)
+ e^{-iq} \qty( 
\hat{\mf b}_q^\dagger \hat{\mf b}_{k+p-q}^\dagger \hat{\mf b}_k  \hat{\mf b}_p 
+ \hat{\mf b}_{k+q+p}^\dagger \hat{\mf b}_k \hat{\mf b}_q  \hat{\mf b}_p
)
] \, . 
\label{eq:ham-long-k}
\end{align}

\subsection*{Collective modes}
Notice that the photon couples to only the $k=0$ mode of the HP boson. 
Therefore, we isolate the $k=0$ mode of the HP boson to investigate the impact of cavity-spin coupling. 
We note that this is the collective mode of interest in the Dicke model, where the $k\neq 0$ modes in \eqref{eq:ham-long-k} do not contribute. 
Thus, the Hamiltonian that controls the superradiant transition in TFIM-Dicke model is expressed in terms of $\hat a$ and $\hat{\mf b}_0$,
\begin{align}
\hat H_0' =&~ \omega_0 \hat a^\dagger \hat a 
 + h \hat{\mf b}_0^\dagger \hat{\mf b}_0
+ \frac{g'}{2} (\hat a + \hat a) (\hat{\mf b}_0 + \hat{\mf b}_0^\dagger) 
- \frac{g'}{(2\sqrt{2SN})^2} (\hat a + \hat a) \qty(\hat{\mf b}_{0}^\dagger \hat{\mf b}_0  \hat{\mf b}_0 
+ \mbox{h.c.} 
)    \\
&~ - \frac{J'}{4}  \qty(\hat{\mf b}_0 \hat{\mf b}_0^\dagger +  \hat{\mf b}_{0} \hat{\mf b}_{0} + \mbox{h.c})+ \frac{J'}{8(2SN)}  \qty[
\hat{\mf b}_0  \hat{\mf b}_{0}^\dagger \hat{\mf b}_0 \hat{\mf b}_0 
+ \hat{\mf b}_0 \hat{\mf b}_0^\dagger \hat{\mf b}_{0}^\dagger \hat{\mf b}_0  
+ \hat{\mf b}_0^\dagger \hat{\mf b}_{0}^\dagger \hat{\mf b}_0  \hat{\mf b}_0 
+ \hat{\mf b}_0^\dagger \hat{\mf b}_0^\dagger \hat{\mf b}_{0}^\dagger \hat{\mf b}_0 + \mbox{h.c.}
] \, . \nn
\end{align}
In order to determine the quadratures that get squeezed, we first determine the classical reference state by substituting
\begin{align}
\hat a \to \expval{\hat a} = \sqrt{2SN} \alpha \, , \qquad
\hat{\mf b}_0  \to \expval{\hat{\mf b}_0} = -\sqrt{2SN} \beta \, ,
\end{align}
with $\alpha, \beta \in \mathbbm{R}$, 
into $H_0'$ to obtain the
 energy density 
\begin{align}
E_0 \equiv \frac{\expval{\hat H_0'}}{2SN} = 
\omega_0 \alpha^2 
+ h \beta^2 
- 2 g' \alpha \beta 
+ g' \alpha \beta^3 
- J' \beta^2 
+ J' \beta^4 \, .
\end{align}

Extremizing $E_0$ leads to 
\begin{align}
&\partial_\alpha E_0 = 0 \Rightarrow 2 \omega_0 \alpha - 2 g' \beta + g' \beta^3 = 0 \, , \\
& \partial_\beta E_0 = 0 \Rightarrow 2 h \beta - 2 g' \alpha + 3 g' \alpha \beta^2 - 2 J' \beta + 4 J' \beta^3 = 0 \, .
\end{align}
These equations are solved to obtain
\begin{align}
& \alpha = \frac{g'}{\omega_0} \beta \qty(1 - \frac{\beta^2}{2})  \, ,\\
& \beta^2 = 
\begin{cases}
    \displaystyle{\frac{1}{2} \frac{g'^2 - g_c'^2}{g'^2 + \omega_0 J'}} & \mbox{if~}  g' > g_c' \\
    0 & \mbox{otherwise}
\end{cases} \, ,
\end{align}
where the critical light-matter coupling
\begin{align}
g_c' = \sqrt{\omega_0 (h-J')} \, .
\end{align}

The effective Hamiltonian governing the fluctuations above this saddle point is obtained from \eqref{eq:ham-long-k} by substituting
\begin{align}
\hat a = \expval{\hat a} + \delta \hat a \qquad \mbox{and} \qquad 
\hat{\mf b}_0 = \expval{\hat{\mf b}_0} + \delta \hat{b} \, .
\end{align}
Thus, up to constant terms, we obtain
\begin{align}
\delta \hat H_0' &= \omega_0 \delta \hat a^\dagger \delta \hat a 
+ \underbrace{\qty[h + 2 g' \alpha \beta - \frac{J'}{4}(2 - 7 \beta^2)]}_{\displaystyle{h_{\text{eff}} }} \delta \hat{b}^\dagger \delta \hat{b} \nn \\
&\quad + \underbrace{\frac{g'}{4} \qty[2 - 3 \beta^2]}_{\displaystyle{g_{\text{eff}} }}(\delta \hat a + \delta \hat a^\dagger) (\delta \hat{b} + \delta \hat{b}^\dagger) \nn \\
&\quad + \underbrace{\frac{1}{4}\qty[2 g' \alpha \beta - J'(1 - 5 \beta^2)]}_{\displaystyle{\Delta_{\text{pair}} }}\qty(\delta \hat{b} \delta \hat{b}  +\delta \hat{b}^\dagger \delta \hat{b}^\dagger ) \, .
\end{align}
This Hamiltonian is diagonalized using the symplectic form $\Sigma = \mqty( \sigma_0 & 0 \\ 0 & -\sigma_0)$, where $\sigma_0$ is the $2\times 2$ identity matrix, to obtain
\begin{align}
\delta \hat H_0' &= \sum_{s = \pm} \eps_s \hat \pi_s^\dagger \hat \pi \, .
\end{align}
The polariton operators, $\hat \pi_s$, are related to the bosonic fluctuation as 
\begin{align}
[\hat \pi_+ \quad \hat \pi_- \quad \hat \pi_+^\dagger \quad \hat \pi_-^\dagger]^T = V  [\delta \hat a \quad \delta \hat b \quad \delta \hat a^\dagger \quad \delta \hat b^\dagger]^T \, ,
\end{align}
where 
\begin{align}
V^{-1} \delta H_0 V = \sigma_0 \otimes  \mqty(\eps_+ & 0\\ 
0 & \eps_- ) \, ,
\end{align}
with $\delta \hat H_0' = [\delta \hat a \quad \delta \hat b \quad \delta \hat a^\dagger \quad \delta \hat b^\dagger]^\dagger ~\delta H_0~ [\delta \hat a \quad \delta \hat b \quad \delta \hat a^\dagger \quad \delta \hat b^\dagger]^T$.

\subsection*{Quadrature and its extremization}
{Here, we identify the quandrature, composed of both photonic and magnonic fluctuations, that support extremal variance. 
Towards this goal, we note that a generic linear combination of $\delta \hat a$ and $\delta \hat b$ takes the form
\begin{align}
\hat d_{z_1, z_2} = z_1 \delta \hat a + z_2 \delta \hat b \, ,
\end{align}
where $z_i \in \mathbbm{C}$. 
We require $\hat d_{z_1, z_2}$ to descibe a bosonic excitation, which implies $[\hat d_{z_1, z_2}, \hat d_{z_1, z_2}^\dagger] = 1$.
This requirement, in turn, constrains 
\begin{align}
|z_1|^2 + |z_2|^2 = 1 \, .
\end{align}
Thus, the most general linear  combination, up to a phase factor, takes the form
\begin{align}
\hat d_{\theta, \psi} = \cos\theta \delta \hat a + e^{i \psi}  \delta \hat b.
\end{align}
We can now construct a generalized quadrature~\cite{meystre2007elements, hayashida2023perfect},  
} 
\begin{align}
\hat X_{\theta, \psi, \phi} &= \half \qty[e^{i\phi} \hat d_{\theta, \psi} + \mbox{h.c.} ] \label{eq:quadra} \\
& = \frac{1}{2} \qty[ 
\qty( 
\cos \theta \delta \hat a + e^{i \psi} \sin\theta \delta \hat b) e^{i\phi} + \mbox{h.c.}  ]\, .
\end{align}
Therefore, the operator $\hat X_{\theta, \psi, \phi}$ is parameterized by the three real-valued angles, $(\theta, \phi, \psi)$, and its variance is given by 
\begin{align}
(\Delta X_{\theta, \psi, \phi})^2 = \expval{\hat X_{\theta, \psi, \phi}^2} - \expval{\hat X_{\theta, \psi, \phi}}^2 \, .
\end{align}
We note that Eq.~\eqref{eq:quadra} is a generalized quadrature for the operator $\hat d_{\theta, \psi}$.

\begin{figure}[!t]
\centering
\includegraphics[width=0.5\linewidth]{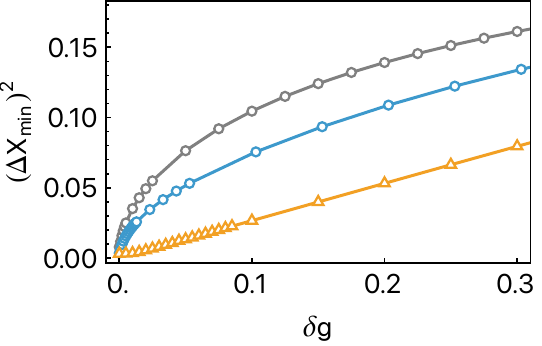}
\caption{{{\bf Minimum variance as a function of $\delta g = g - g_c$ at $h= h_\mathrm{TFIM}$ (orange), $h>h_\mathrm{TFIM}$ (blue), and in the Dicke model (gray).} 
The plot markers indicate numerically computed data points, while the solid curves are
guides for the eye. The saturation-like behavior of the orange data-set is due to finite size effects. 
We note the general suppression of $(\Delta X_\mathrm{min})^2$ as $h \to h_\mathrm{TFIM}$. Here and in the corresponding figure of the main text, we have fixed $(\omega_0, J, N_a, N)= (1, 1, 40, 100)$ where $N_a$ is the cutoff of the photon's Hilbert space dimension. For the Dicke model we have set $(\omega_0, h, J) = (1, 1, 0)$ with the same $N_a$ and $N$.}
}
\label{fig:delta-g}
\end{figure}

Numerically minimizing the variance identifies an optimal choice of the three parameters, and the behavior of the minimized variance is plotted in the main text. 
{In Fig.~\ref{fig:delta-g} we plot the dependence of the polaritonic quadrature that supports the least variance on $\delta g = g - g_c(h)$ at the same values of $h$ as in Fig.~2(e).
}

\subsection*{Quantum Fisher Information}
Here, we will recall a well-known result in quantum estimation theory, for pure states  $F_Q(\hat X) = 4 \Delta \hat X^2$. 
For more general results along the same direction, the reader is directed to Ref.~\cite{paris2009quantum}.
We begin by considering a phase estimation problem, where we want to know how precisely a parameter, $\vtheta$, can be estimated by $M$ measurements on a pure  state,
\begin{align}
\ket{\psi(\vtheta)} = e^{- i \vtheta \hat X} \ket{\psi_0}
\label{eq:psi}
\end{align}
with $\ket{\psi_0}$ being an initial pure state which helps generate a family of pure states through the action of the unitary operator $e^{- i \vtheta \hat X}$ (therefore, $\hat X$ is Hermitian).
We note that Eq.~\eqref{eq:psi} defines a unitary encoding of the parameter $\vtheta$.
Without loss of generality, we will assume $\braket{\psi_0}{\psi_0} = 1$ such that the density matrix $\rho(\vtheta) \equiv \ket{\psi(\vtheta)}\bra{\psi(\vtheta)} = \rho(\vtheta)^2$.
It is an interesting fact that the phase estimation problem crucially depends on the local geometry of the Hilbert space to which $\ket{\psi(\vtheta)}$ belongs.
In particular, tuning $\vtheta$ generates a trajectory in the Hilbert space, and the minimum possible variance of $\vtheta$ controls the distinguishability between two ``neighboring'' states on this trajectory or manifold~\cite{braunstein1994statistical}. We will see this connection explicitly. 

The quantum Fisher information (QFI) can be introduced as the upper bound on Fisher information~\cite{braunstein1994statistical,paris2009quantum} , $F(\vtheta) \leq F_Q(\vtheta)$ with
\begin{align}
 F_Q(\vtheta) = \tr{\partial_\vtheta \rho(\vtheta) L(\vtheta)}.
\end{align}
Here, $L(\vtheta)$ is the symmetric logarithmic derivative, which obtains a simple form for pure states~\cite{paris2009quantum} 
\begin{align}
L(\vtheta) = 2 \partial_\vtheta \rho(\vtheta) = \ket{\partial_\vtheta \psi(\vtheta)}\bra{ \psi(\vtheta)} + \ket{\psi(\vtheta)}\bra{\partial_\vtheta \psi(\vtheta)}.
\end{align}
Owing to our definition of $\ket{\psi(\vtheta)}$, $F_Q(\vtheta)$ implicitly depends on $\hat X$, and we obtain
\begin{align}
F_Q(\vtheta) &= 4 \qty[ \expval{\hat X^2}{\psi(\vtheta)} - \expval{\hat X}{\psi(\vtheta)}^2 ] \\ 
&= 4\qty[ \expval{\hat X^2}{\psi_0} - \expval{\hat X}{\psi_0}^2 ] \\
&= 4 \Delta \hat X^2,
\end{align}
which is the aforementioned 
%desired 
relationship.
In the `Methods', we have exchanged   $\vtheta$  with $\hat X$  in the argument of $F_Q$ to explicitly refer to its dependence on the choice of $\hat X$. 

Viewed from the perspective of distinguishability in the Hilbert space~\cite{braunstein1994statistical}, we now demonstrate how the variance of $\hat X$ controls the Fubini-Study metric along the one-dimensional manifold parameterized by $\vtheta$.
The metric is obtained from the overlap
\begin{align}
|\braket{\psi(\vtheta)}{\psi(\vtheta + \delta \vtheta)}|^2 = 1 + 2\Re{\braket{\psi(\vtheta)}{\partial_\vtheta \psi(\vtheta)}} \delta \vtheta 
+ \qty[|\braket{\psi(\vtheta)}{\partial_\vtheta \psi(\vtheta)}|^2],
\end{align}
where $\ket{\partial_\vtheta \psi(\vtheta)} \equiv \partial_\vtheta \ket{\psi(\vtheta)} = - i \hat X \ket{\psi(\vtheta)}$.
For our case, we take advantage of the fact that $\ket{\psi(\vtheta + \delta \vtheta)} = \qty[1 - i \langle\hat X \rangle \dl \vtheta - (1/2) \langle\hat X^2 \rangle \dl \vtheta^2 + \order{\delta \vtheta^3}]$ to deduce that the infinitesimal distance on the manifold 
\begin{align}
ds^2 &\equiv 1 - |\braket{\psi(\vtheta)}{\psi(\vtheta + \delta \vtheta)}|^2 \\
&= \Delta \hat X^2 \delta \vtheta^2 + \order{\delta \vtheta^3}.
\end{align}
Thus, the QFI quantifies the speed of state-evolution along the manifold,
\begin{align}
 \lim_{\delta \theta \to 0} \frac{ds}{\delta \vtheta} = \frac{1}{2} \sqrt{F_Q(\vtheta)}.
\end{align}
For $\hat X = \hat X_{\text{max}}$ the variance diverges as the TFIM QCP is approached, and so does the QFI. 
Therefore, in the vicinity of the TFIM QCP, the states on the manifold generated by $\hat X_{\text{max}}$ become highly resolved.

\section{\label{app:non-crit} Exact solution of cavity coupling to a non-critical mode of TFIM}
In this section we give a complete derivation on the exact solution of the cavity coupled to a non-critical mode of TFIM.
While the general expression of the free energy has been obtained earlier~\cite{gammelmark2011phase}, the characterization of the tricritical point, including the scaling form in Eq~(10) and Fig.~3(b) in the main section, represents original results of the present work.

By introducing the coherent photon basis to the partition function, we have
\begin{align}
    Z=\int D[a^{\dagger},a] e^{-\beta \omega_{0}a^{\dagger}a} \text{Tr}_{spin}e^{-\beta H_{a}} \, , 
    \label{partition}
\end{align}
where $H_{a}=H_{\text{spin}}+\frac{g}{\sqrt{N}}(a+a^{\dagger})\sum_{\mathbf{r}}S_{\mathbf{r}}^{y}$. 
The TFIM with non-singular light-spin coupling $\frac{g}{\sqrt{N}}(a+a^{\dagger})\sum_{\mathbf{r}}S_{\mathbf{r}}^{y}$ in the coherent photon basis can be diagonalized exactly via Jordan-Wigner transformations. 

Rotating the system around $x$ axis, such that new $\hat{S}_z$ couples with effective field $h_{\text{eff}}=\sqrt{h^2+g^{2}(a+a^{\dagger})^{2}/N}$: $-h_{\text{eff}}\sum_{\mathbf{r}}\hat{S}^{z}_{\mathbf{r}}$.
We diagonalize it using Jordan-Wigner transformation,
\begin{align}
    S^{z}_{\mathbf{r}}&=1/2-c^{\dagger}_{\mathbf{r}}c_{\mathbf{r}} \,  \\
    S^{x}_{\mathbf{r}}&=\frac{1}{2}\prod_{r'<r}e^{i\pi c^{\dagger}_{\mathbf{r'}}c_{\mathbf{r'}}}(c^{\dagger}_{\mathbf{r}}+c_{\mathbf{r}}) \, .
\end{align}

The Hamiltonian becomes:
\begin{align}
    H_{a}=-\frac{J}{4}\sum_{\langle \mathbf{r},\mathbf{r}'\rangle}(c^{\dagger}_{\mathbf{r}}-c_{\mathbf{r}})(c^{\dagger}_{\mathbf{r}'}+c_{\mathbf{r}'})-h_{\text{eff}}{\sum_{\mathbf{r}}(1/2-c^{\dagger}_{\mathbf{r}}c_{\mathbf{r}})} \, .
\end{align}

After Fourier transformation:
$c_{k}=\frac{1}{\sqrt{N}}\sum_{\mathbf{r}}c_{\mathbf{r}}e^{ikr}$, we get $H=\sum_{k}\Psi^{\dagger}_{k}H_{k}\Psi_{k}$
where $\Psi^{\dagger}_{k}=(c^{\dagger}_{k},\, c_{-k})$, and
\begin{equation}
    H_{k}=\frac{1}{4}\begin{pmatrix}
  2h_{\text{eff}}-J\cos{k} & iJ\sin{k} \\
  -iJ\sin{k} & -(2h_{\text{eff}}-J\cos{k})
\end{pmatrix}
\, .
\end{equation}

\begin{figure}[!t]
\centering
\includegraphics[width=0.6\columnwidth]{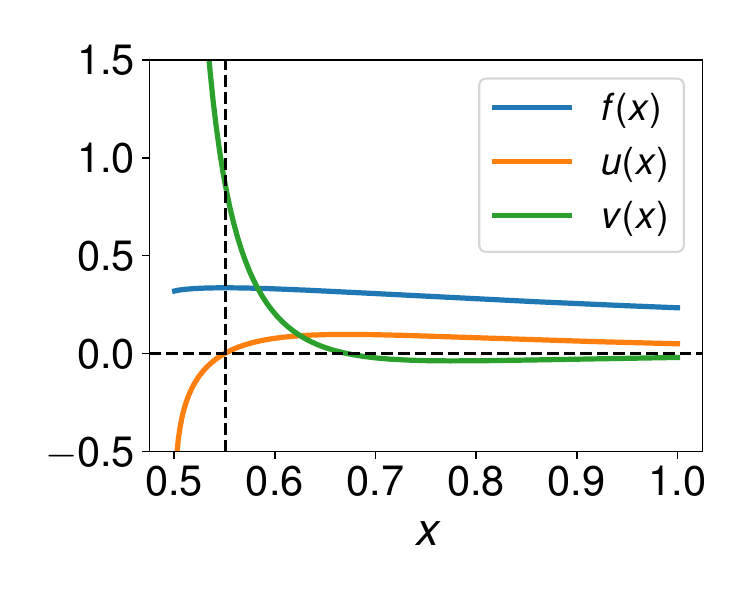}%
\caption{\textbf{Expansion coefficients of the Landau free energy.} 
Dependence of the quadratic $f(x)$, quartic $u(x)$, and sixth-order $v(x)$ coefficients on the scaled magnetic field $x=h/J$. 
The vertical dashed line marks the tricritical point at $h/J \approx 0.55$, separating first-order and second-order transitions.}
\label{fig.landau}
\end{figure}

The free energy density at zero temperature has the form after integrating out the fermions:\begin{align}
 \mc E_{g}(\phi) = \frac{1}{4}\omega_{0}\phi^{2}
 -\frac{1}{2\pi}(J+2h_{eff})E\left[\frac{8Jh_{eff}}{(J+2h_{eff})^2}\right]  \, ,   
\end{align}
where $h_{eff}=\sqrt{h^2 + g^2 \phi^2}$ is the effective transverse field of the TFIM. $E(x)$ is the complete elliptic integral of the second kind. $\phi=\langle a+a^{\dagger}\rangle/\sqrt{N}$ is the order parameter for superradiant phase. 
In the field-polarized phase ($h/J>0.5$), the free energy density can be expanded up to sixth order in $\phi$:
\begin{align}
\frac{\mc E_{g}(\phi)}{J} = \qty(\frac{\omega_{0}}{4J} -f(h/J)\frac{g^2}{J^{2}}){\phi}^{2} 
+ u(h/J)\frac{g^4}{J^4} \phi^{4} 
+ v(h/J) \frac{g^6}{J^6} \phi^{6} \, ,
\end{align}
where
\begin{align}
f(x)&= \frac{1}{2\pi}\left\{\frac{\mathrm{E}\left(\frac{8 x}{(1 + 2 x)^2}\right)}{x} 
+ \frac{(-1 + 2 x)\, \pi\, 
{}_2F_1\left(\frac{1}{2}, \frac{3}{2}; 2; \frac{8 x}{(1 + 2 x)^2}\right)}
{2 x (1 + 2 x)^2}\right\} \, ,  \\
u(x)&= \frac{  (1 + 2x)^2\, \mathrm{E}\left( \frac{8x}{(1 + 2x)^2} \right) 
- \mathrm{K}\left( \frac{8x}{(1 + 2x)^2} \right) } 
{16\pi x^4 \left( 1 + 2x  \right)}  \, , \\
v(x)&= 
\frac{
-(1 + 2x)(-2 + 7x^2)\, \mathrm{E}\left( \frac{8x}{(1 + 2x)^2} \right)
+ (-1 + 2x)(2 + x^2)\, \mathrm{K}\left( \frac{8x}{(1 + 2x)^2} \right)
}{
48 \pi x^6 ( -1 + 4x^2 )
} \, ,
\end{align}
where $_2F_{1}(a,b,c;z)$ is the hypergeometric function. $K(x)$ is the complete elliptic integral of the first kind. The three functions are plotted in Fig.\ref{fig.landau}.

In the vicinity of the tricritical point,  $f(h/J)=0.336+O(\delta h^2)$, $u(h/J)=  2.9\delta h+O(\delta h^{2})$, $v(h/J)= 0.88+O(\delta h)$ with $\delta h=(h-h_{tri})/J$ with $h_{tri}\approx 0.55J$, and the order parameter has the form 
\begin{align}
\mc N = \phi^2 =\frac{J^{2}}{4g^{2}}\frac{\delta g}{u+\sqrt{u^2+\frac{3}{4}v\delta g}}\theta(\delta g) \, .
\end{align}

% \section*{Section V: Cavity coupling to the anisotropic XY model}
\section{Cavity coupling to the anisotropic XY model}

In this section we consider  cavity coupling to an anisotropic XY model, described by the Hamiltonian
\begin{align}
    H = -\frac{J}{2}\sum_{\langle i,j \rangle} 
    \big[(1+\Delta)S_{i}^{x}S_{j}^{x} + (1-\Delta)S_{i}^{y}S_{j}^{y}\big]
    - \frac{g}{\sqrt{N}}(a+a^\dagger)\sum_{i}\mathbf{n}\cdot\mathbf{S}_{i} \, .
\end{align}

Because the $\hat{x}$ and $\hat{y}$ directions are equivalent up to the sign of $\Delta$, it is sufficient to distinguish two qualitatively different cases depending on whether the cavity couples to a {singular} magnetic order parameter or to a {non-singular} operator of the spin system.  

Case I: Coupling to a singular mode ($\mathbf{n}=\hat{x}$). 
In this case, the cavity photons couple directly to the ferromagnetic order parameter of the XY chain.  
At $\Delta=-1$, the model reduces to the Dicke–TFIM at $(h,\mathbf{n})=(0,\hat{y})$. For $g \ll \sqrt{J\omega_{0}}$, the ground state is a ferromagnetic state polarized along $S^{y}$, while for $g \gg \sqrt{J\omega_{0}}$ it becomes a superradiant state. The superradiant quantum phase transition (SRPT) between these two states is known to be discontinuous.
However, in the regime $0 \geq \Delta > -1$, correlations exist in both the $x$ and $y$ spin components.  
As $\Delta \to 0^{-}$, approaching the XY {QCP}, the correlation length of $\hat{S}^{x}$ fluctuations diverges, driving the system quantum critical. This divergence continuously suppresses the critical coupling $g_{c}$ for the SRPT, even though the
%magnetic 
{matter} sector remains ordered along $\hat{S}^{y}$.

DMRG simulations confirm this expectation: direct cavity coupling to $\hat{M}_{x}$ strongly reduces $g_{c}$ below the ultrastrong-coupling regime as $\Delta$ is tuned from $-1$ to $0$ [see Fig.~\ref{fig:app-xy}{\bf c}].

Case II: Coupling to a non-singular mode ($\mathbf{n}=\hat{z}$).
Here, the cavity mode couples to $\hat{S}_{z}$, which is not an order parameter of the XY model. In this case, the model can be solved exactly. Applying the Jordan–Wigner transformation, the Hamiltonian is mapped to a quadratic fermionic form,
\begin{align}
    H = \sum_{k} \Psi^{\dagger}_{k} H_{k} \Psi_{k} \, , \qquad 
    \Psi^{\dagger}_{k} = (c^{\dagger}_{k},\, c_{-k}) \, ,
\end{align}
with
\begin{equation}
    H_{k} =
    \frac{1}{4}\begin{pmatrix}
        2g\phi - J\cos{k} & iJ\Delta\sin{k} \\
        -iJ\Delta\sin{k} & -(2g\phi - J\cos{k})
    \end{pmatrix} \, .
\end{equation}
Integrating out the fermions yields the zero-temperature free energy density
\begin{align}
    \frac{\mathcal{E}_{g}(\phi)}{N} 
    = \frac{1}{4}\,\omega_{0}\phi^{2}
    - \frac{1}{8}\int_{-\pi}^{\pi}\frac{dk}{2\pi}\,
      \sqrt{(2g\phi - J\cos{k})^{2} + \Delta^{2}J^{2}\sin^{2}{k}}\, .\label{xyenergy}
\end{align}
The free energy derived above provides the theoretical framework for this case.  
Analytical calculations  show that the SRPT is always discontinuous, including at the magnetic {QCP} $\Delta=0$, as is summarized in Fig.~\ref{fig:app-xy}({\bf a,b}).  
This reflects the non-singular nature of $\hat{M}_{z}$ in the spin system.  
Nevertheless, the critical coupling is suppressed at $\Delta=0$, reaching $g_{c}\approx0.78\sqrt{\omega_{0}J}$, compared with $g_{c}\approx0.915\sqrt{\omega_{0}J}$ in the Ising limit $\Delta=\pm1$.  

In both cases, the critical coupling $g_{c}$ decreases as the XY chain approaches its magnetic {QCP} at $\Delta=0$.  
This demonstrates that magnetic quantum fluctuations generally enhance the onset of the superradiant phase, independent of whether the cavity couples to a singular or non-singular operator of the spin system.

\begin{figure}[!t]
\centering
\includegraphics[width=0.9\linewidth]{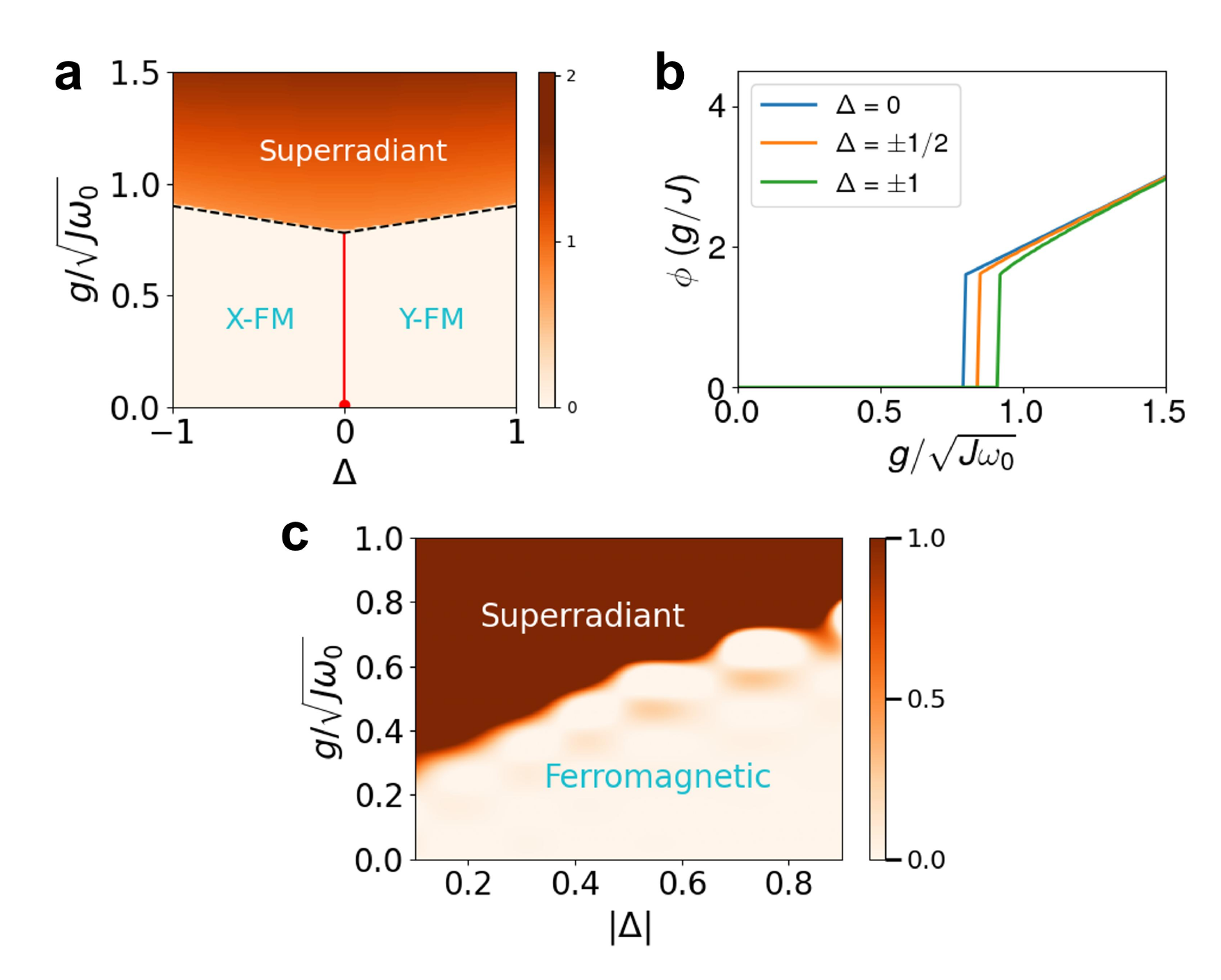}
\caption{{\bf Quantum phase transitions in the anisotropic XY model coupled to cavity photons. ~~~}{\bf a}, Zero-temperature phase diagram of the anisotropic XY model coupled to cavity photons for $\mathbf n = \hat z$.  
The vertical solid line at $\Delta=0$  marks the QCP of the anisotropic XY model where the excitations beomes gapless. 
Dashed lines  denote the first order transition from ferromagnetically ordered normal phase to the superradiant phase. 
{\bf b}, Superradiant order parameter $\phi=\langle a+a^{\dagger} \rangle/\sqrt{N}$ with respect to light-matter coupling $g$. {{\bf c}, Zero-temperature phase diagram of the anisotropic XY model coupled to cavity photons for $\mathbf n = \hat x$ [cf. Eq.~(1)] and $\Delta < 0$ [cf. Eq.~(4)]  as  the $XY$ {QCP} at $\Delta = 0$ is approached.} 
The color bar indicates the value of  $\widetilde \Theta(\mc N) = \Theta(\mc N - 1) + \mc N \Theta(1 - \mc N)$, where  $\Theta$ is  the Heaviside theta function and $\mc N$ is obtained from DMRG simulations.
}
\label{fig:app-xy}
\end{figure}

{
\section{Practical and formal aspects of cavity coupled quantum magnets} \label{app:additional}
In this section we discuss details of  potential cavity designs suitable for the setting discussed here, the formal prescription of the thermodynamic limit,  additional details on the absence of additional $\mathbf A^2$ dependent terms, and the path towards practical applications of the models discussed in this work.

\subsection*{Cavity geometry and thermodynamic limit}
In this section we discuss a sample cavity geometry for the cavity-TFIM setup discussed in this manuscript, and utilize it to define the thermodynamic limit.
We consider a cuboid cavity with linear dimensions $L_x$, $L_y$, and $L_z$ that are constrained as $L_x \gg L_y, L_z$.
Perfectly reflective mirrors are assumed to be present on the surface terminations along $\hat y$ and $\hat z$ (see schematic Fig. ABC).
This arrangement generates the quantized electromagnetic modes within the cavity~\cite{kakazu1994quantization, roman2025bound}.
The collective coupling controlling the Zeeman interaction between the magnetic field within the cavity and the quantum spins within the magnetic material depends on the cavity and material dimensions as
\begin{align}
g \propto \sqrt{\frac{a_x a_y a_z N_x N_y N_z}{L_x L_y L_z}},
\end{align}
where $N_j$ ($a_j$) is the number of spins (lattice spacing) along the $j$-th direction.

For an effective 1D Ising material, such as CoNb$_2$O$_6$, multiple Ising chains are present that interact weakly with each other. 
By orienting the chains along $\hat x$, such that each chain is of length  $L_\mathrm{chain} = a_x N_x$, we assume $N_y, N_z \sim \order{1}$ and ignore the inter-chain interactions to obtain an effective description of the system in terms of the TFIM. 
$L_y$ and $L_z$ control the separation between each pair of mirrors and are assumed to be fixed. 
The thermodynamic limit is taken by sending $L_x \to \infty$ at a fixed ratio  $L_\mathrm{chain} / L_x$.
This process keeps both the density of spins within the cavity and $g$ to be finite.

We note that our results also apply to 2D magnetic materials, such as CrI$_3$ which is a van der Waals Ising ferromagnet.
The cavity design for 2D materials is less restrictive as only a single pair of mirror (say along $\hat z$) suffices, and the thermodynamic limit is obtained by sending the transverse area of the cavity $L_x L_y \to \infty$ with $(a_x a_y N_x N_y)/(L_x L_y)$ held fixed.

\subsection*{A microscopic justification for gauge invariance of the minimal model}
In general both   electric and magnetic interactions between the cavity photons and the matter modes are  present at microscopic scales. 
To demonstrate it, we consider a Hubbard model coupled to cavity-photons.
The Hubbard model is assumed to describe nearest-neighbor hoppings and an on-site Coulomb interaction, whose interplay generates the effective spin-spin interaction modeled by the Ising model studied in our paper.  
The microscopic gauge invariant Hamiltonian takes the form
\begin{align}
H &= -t \sum_{\langle ij\rangle} \sum_{s} \exp{-i \frac{e}{\hbar} \mathbf A\qty(\frac{\mathbf r_i + \mathbf r_j}{2}) \cdot (\mathbf r_i - \mathbf r_j)} c_{s,i}^\dagger c_{s, j} + U \sum_{j}   n_{\uparrow, j} n_{\downarrow, j} \nn \\
& \quad + g_L \mu_B \sum_{j, s, s'} \mathbf B_j \cdot c^\dagger_{s, j} \mathbf \sigma^{ss'} c_{s', j},
\label{eq:full-H}
\end{align}
where $\mathbf r_j$ is the postion of the $j$-th lattice site.
The final term is the microscopic Zeeman interaction with the cavity-magnetic field $\mathbf B_j = \mathbf B(\mathbf r_j) = \grad \cross \mathbf A(\mathbf r_j)$.
Since it is manifestly gauge-invariant, it does not produce additional terms. 

In the $t \ll U$ limit and in the absence of the Zeeman term, Sentef {et al.}~\cite{sentef2020quantum}, working within the single-mode approximation, showed that the superexchange interaction, $J$, obtained within the present setting acquires a dependence on the cavity-photons: $J \to J(\hat a, \hat a^\dagger)$.
The photon dressing of $J$, however, is controlled by the dimensionless coupling appearing in the lattice version of Pierls substitution: $t_{ij} \to t_{ij} \exp{- i g_P (\hat a + \hat a^\dagger} $, where in our present convention $g_P = \sqrt{2} e a_\mathrm{chain}/\sqrt{\hbar L_x L_y L_z \epsilon_0 \omega_0}$ with $\epsilon_0$ being the permittivity within the cavity.
In the limit $g_P \ll 1$ (an $10$ GHz cuboid microwave cavity with volume $10$ cm$^3$ and $a_\mathrm{chain} = 0.5$ nm leads to a $g_P \sim 10^{-9}$), $J(\hat a, \hat a^\dagger) = J(0) + \order{g_P^2 \hat a^2, g_P^2 (\hat a^\dagger)^2}$.
Thus, the leading  order term is  independent of $\hat a$ and $\hat a^\dagger$.
Approaching the SRPT from the normal phase ($\expval{\hat a} = 0$), we note that the sub-leading terms are not only numerically suppressed but also result in interaction vertices with higher number of operators compared to those already included in our model (none of which involve spatial derivatives) which makes them irrelevant in a scaling sense.
Moreover, this procedure does not generate any renormalization to the Zeeman term.
Therefore, our model accounts for all leading order effects of the cavity field in the strong coupling limit of the Hubbard model ($t \ll U$). 
The effects of additional cavity-mode mediated vertices is suppressed, and they are not expected to alter the conclusions of our work.

A physical justification for this outcome is the fact that terms like $A^2 c^\dagger_i c_j $  that raises questions like a potential no-go theorem originate from hopping of electrons.
While such diamagnetic terms are  important for the physics of nearly free electrons within a cavity due to their high mobility, as $U$ increases, the electrons tend to localize, and the diamagnetic contribution would be  expected to be suppressed. 
In the regime where the effective description of the system is in terms of a quantum magnet, individual electrons can no longer hop over long distances, and the dominant dynamics is that of charge-neutral spin-fluctuations which do not directly couple to the electromagnetic gauge field, and, thus, will be unaffected by the electric component of the gauge field.

We note that the general conclusion in this section that strong electronic correlations suppresses the role of the diamagnetic term in determining the presence of an SRPT is supported by gauge invariant analysis of the cavity-coupled Hubbard model in the context of excitonic instabilities~\cite{mazza2019superradiant}.
In the case of cavity coupled to a critical degree of freedom, at a fixed $g$, the phase boundary separating the normal state from the superradiant ferromagnetic state can alternatively be interpreted as a persistence of the ferromagnetic order in the TFIM beyond its usual (i.e. free space) confines, $h < h_\mathrm{TFIM}$.
This perspective is also aligned with the findings in Ref.~\cite{mazza2019superradiant}.

\subsection*{Multi-mode cavity-coupled quantum magnets}
Although in phenomenological models of experiments in cavity magnonic systems, it usually suffices to  truncate the number of cavity modes ($M$) to a small number, even $M=1$~\cite{},  
the single-mode cavity  is still an idealization, since realistic cavities generally possess multiple modes that interact with the matter degrees of freedom. 
In this section we explore the effects of multiple cavity modes on the key conclusion in this work: quantum criticality in the matter sector strongly renormalizes the condition for and nature of SRPTs and, with a suitable choice of light-matter coupling, can even dramatically favor the formation of superradiant states.

Since quantum critical phenomena in quantum materials are formally defined in the thermodynamic limit, it would be expected that that the relevant dimensions of the material must be a substantial fraction, if not comparable, to a subset of the cavity's dimensions in order to facilitate substantial hybridization between the (near-)critical matter modes with the cavity modes.
This requirement naturally leads to a local description of the interaction between the two  degrees of freedom, which involves multiple cavity modes, especially those that are labeled by the momentum components parallel to the thermodynamically relevant dimensions of the quantum materials. 
Thus, for an 1D Ising magnet, the most relevant terms in the Hamiltonian controlling the low-energy physics  within our framework is 
\begin{align}
H_\mathrm{multi}^{\mathrm{gen}} = \sum_{n_{x}} \omega_{n_x} a_{n_x}^\dagger a_{n_x} + \sum_{i = 1}^{N}  B_{i}  S^x_{i} + H_\mathrm{spin} - h \sum_{i = 1}^{N} S^z_{i},
\label{eq:H-all}
\end{align}
where $n_{x}$ is the mode index of the momentum $k=\frac{2\pi n_x}{L_x}$ along $L_x$, $N$ is the number of spins, and $B_i$ is the local cavity-magnetic field (following the setup discussed in Sec.~\ref{app:additional})A this is parallel to the Ising chain direction), $N$ is the number of spins, $B_i$ is the local cavity-magnetic field, and $H_\mathrm{spin}$ controls the spin-spin interactions in the 1D magnet.
When the fixed ratio $L_\textrm{chain}/L_x \ll 1$, the cavity-magnetic field can be considered as  uniform within the sample, which will result in the following multi-mode generalization of the single-mode model studied in our paper,
\begin{align}
H_\mathrm{multi}' = \sum_{n_x=1}^M \omega_{n_x} a_{n_x}^\dagger a_{n_x} - h S^z_T + \frac{i}{\sqrt{N}} \sum_{n_x=1}^M g_{n_x} (a_{n_x} - a_{n_x}^\dagger) S^x_T + H_\mathrm{spin}.
\label{eq:multimode}
\end{align}
where $S^\mu_T = \sum_{j=1}^N S^\mu_j$ and we have assumed the dissimilar sizes of the cavity and the sample allows independent tuning of $M$ and $N$.
We note that this model is a direct generalization of the multi-mode Dicke model~\cite{tolkunov2007quantum} to our setting.

We follow the path integral method utilized in Sec ~\ref{app:GF}. By integrating out the photon degrees of freedom, we obtain the effective spin susceptibility $\chi(t)\equiv-\frac{i}{N}\langle \mathcal{T}S^{x}_{T}(t)S^{x}_{T}(0)\rangle$ for the total spin $S_{T}^{x}$. For $H_\textrm{spin} = - (J/N) (S^x_T)^2$, in frequency space, this yields:
\begin{align}
    \chi(\omega)=\frac{2h}{\omega^2-h\left(h-J+\sum_{n_x}\frac{g_{n_x}^2\omega_{n_{x}}}{\omega^2-\omega_{n_x}^2}\right)}.
\end{align}

The SRPT occurs when the static spin susceptibility diverges at finite light-matter coupling, namely: $\chi(\omega\rightarrow0)_{h=h_c}\rightarrow\infty$. This divergence establishes that the multi-mode system $H_{\mathrm{multi}}$ continues to support an SRPT with an onset at: $h_{c}=J+\sum_{n_x=1}^M g_{n_x}^2/\omega_{n_x}$. Our result is also consistent with a suitable generalization of the results in Ref.~\cite{tolkunov2007quantum}.
Although $h_c$ is dependent on the mode profile ($\omega_k$) and the non-uniform light-matter coupling $g_k$, we observe that as $(h-J) \to 0^+$, our statement on the proximity to the magnetic QCP facilitating SRPTs at progressively weaker light-matter coupling continues to hold.
We also note that although a systematic single-mode limit does not appear to be available in any physically realistic limits of $H_\mathrm{multi}$, the present analysis implies that the qualitative outcome in the physically attainable $L_\textrm{chain}/L_x \ll 1$ limit (relevant to typical cavity-magnonics setups) was  anticipated by our single-mode analysis,  underscoring the experimental relevance of our results.

}

\end{document}